\documentclass[preprint, 11pt]{elsarticle}
\makeatletter
\def\ps@pprintTitle{%
	\let\@oddhead\@empty
	\let\@evenhead\@empty
	\def\@oddfoot{\centerline{\thepage}}%
	\let\@evenfoot\@oddfoot}
\makeatother

\usepackage[utf8]{inputenc}
\usepackage{graphicx}
\usepackage{color}
\usepackage{booktabs}
\usepackage{amsmath,amsthm,amssymb,amsfonts,amscd,amsbsy,amsxtra}
\usepackage{wasysym}
\usepackage{ae} 
\usepackage{pgf}
\usepackage[labelformat=simple]{subcaption}

\usepackage {ulem}
\usepackage[pagewise]{lineno}
\usepackage{upgreek}
\usepackage{arydshln}
\usepackage{units}
\usepackage{hyphenat}
\usepackage{enumitem}
\usepackage{multirow}

\RequirePackage[colorlinks,citecolor=blue,urlcolor=blue]{hyperref}

\newcommand{\ind}{{1\hspace{-1mm}{\rm I}}}
\newcommand{\E}{\mathbb{E}}

\newcommand{\R}{\mathbb{R}}

\newcommand{\N}{\mathbb{N}}
\newcommand{\C}{\mathbb{C}}

\begin{document}
	
\begin{frontmatter}
\title{Stochastic model for the 3D microstructure of pristine and cyclically aged cathodes in Li-ion batteries}


\author[1]{Klaus Kuchler \corref{cor1}}
\author[1]{Daniel Westhoff}
\author[1]{Julian Feinauer}
\author[1,2]{Tim Mitsch}
\author[2]{Ingo Manke}
\author[1]{Volker Schmidt}

\address[1]{Institute of Stochastics, Ulm University, 89069 Ulm, Germany}
\address[2]{Institute of Applied Materials, Helmholtz-Zentrum Berlin, 14109 Berlin, Germany}

\cortext[cor1]{Corresponding author. Email: klaus.kuchler@uni-ulm.de. Phone: +49 731 50 - 23555. Fax: +49 731 50 - 23649.}

\begin{abstract}
It is well-known that the microstructure of electrodes in lithium-ion batteries strongly affects their performance. Vice versa, the microstructure can exhibit strong changes during the usage of the battery due to aging effects. For a better understanding of these effects, mathematical analysis and modeling has turned out to be of great help. In particular, stochastic 3D microstructure models have proven to be a powerful and very flexible tool to generate various kinds of particle-based structures. Recently, such models have been proposed for the microstructure of anodes in lithium-ion energy and power cells. In the present paper, we describe a stochastic modeling approach for the 3D microstructure of cathodes in a lithium-ion energy cell, which differs significantly from the one observed in anodes. The model for the cathode data enhances the ideas of the anode models, which have been developed so far. It is calibrated using 3D tomographic image data from pristine as well as two aged cathodes. A validation based on morphological image characteristics shows that the model is able to realistically describe both, the microstructure of pristine and aged cathodes. Thus, we conclude that the model is suitable to generate virtual, but realistic microstructures of lithium-ion cathodes.
\end{abstract}


\end{frontmatter}

%\linenumbers
%\modulolinenumbers[5]

\section{Introduction}\label{sec:intro}

One of the most common energy storage devices are lithium-ion (Li-ion for short) batteries which have a wide range of applications including automotive technology for electromobility. Thus, there is a variety of publications on Li-ion batteries, where issues like performance, storage capacity, durability and aging effects of Li-ion batteries are discussed, see, e.g.,~\cite{barre.2013, broussely.2005, choi.2012, etacheri.2011}.

It is well-known that the above mentioned issues are more or less directly linked to the morphology of battery electrodes, see, e.g.,~\cite{cho.2015, garcia.2005, wang.2017}. Especially for positive electrodes, i.e., cathodes, microstructure-property relationships were considered in~\cite{wiedemann.2013}. To get a meaningful and valid insight into microstructure-property relationships of Li-ion batteries, it is highly relevant that one has detailed information of the 3D electrode morphology. For a long time, this maybe was one of the main limitations, since there were no advanced (imaging) techniques to measure the highly-complex 3D microstructure of electrodes. Furthermore, also most modeling approaches which were developed to describe electrochemical processes as transport behavior of Li-ions, reactions on the surface of active material or degradation and aging phenomena have been based on 1D or 2D considerations, see, e.g.,~\cite{bower.2011, newman.1975}. But by the use of modern imaging techniques~\cite{eastwoodETAL.2014, mitsch.2014, shearing.2011} combined with larger storage and computation capacities on computers, it is nowadays possible to reconstruct and characterize the 3D microstructure of Li-ion batteries. On such three-dimensional structures, spatially resolved numerical modeling approaches can be used to describe and predict electrochemical processes, see, e.g.,~\cite{latz.2011, latz.2011b, yan.2012}. Thus, the combination of 3D microstructures as input for spatially resolved simulation of electrochemical processes is a powerful tool to investigate the influences of morphologies on batteries' functionality. One way to provide these 3D microstructures is to directly take data which was measured and reconstructed from tomographic imaging, but a more efficient and more flexible way is to use stochastic 3D microstructure modeling, which has already been successfully applied for various kinds of materials, see, e.g.,~\cite{gaiselmann.2012, stenzel.2013, thiedmann.2011}. This has the following advantages.
To calibrate the stochastic microstructure model just a few (or maybe one) representative tomographic images are required which saves repeated expensive tomographic measurements. Furthermore, any number of desired statistically equivalent replications of a morphology can be generated without using laboratory resources every time. But a key advantage is the opportunity to virtually generate new and not yet manufactured materials with morphologies leading to a desirable functionality of the battery. This idea is called virtual materials testing and was already successfully performed, e.g., in~\cite{gaiselmann.2014}.

\begin{figure}[!ht]
	\centering
	\begin{subfigure}[c]{0.32\textwidth}
		\includegraphics[width=\textwidth]{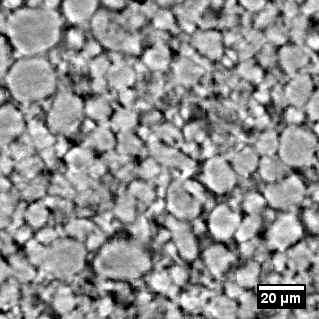}
		\caption{Pristine anode in~\cite{feinauer.2015b}}
		\label{subfig:anodeEnergy}
	\end{subfigure}
	\begin{subfigure}[c]{0.32\textwidth}
		\includegraphics[width=\textwidth]{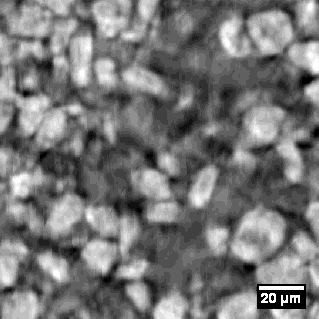}
		\caption{Pristine anode in~\cite{westhoff.2017}}
		\label{subfig:anodePower}
	\end{subfigure}
	\begin{subfigure}[c]{0.32\textwidth}
		\includegraphics[width=\textwidth]{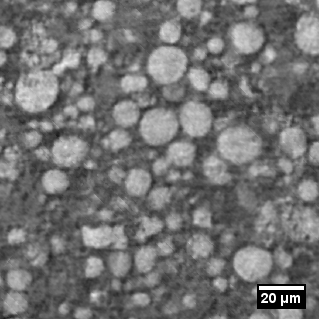}
		\caption{Pristine cathode}
		\label{subfig:cathodeEnergy}
	\end{subfigure}
	\caption{Tomographic grayscale images - 2D slices of cutouts - from anodes considered in~\cite{feinauer.2015b} and~\cite{westhoff.2017} compared to a pristine cathode considered in the present paper.}
	\label{fig:compAnodesCathode}
\end{figure}
In the present paper, we enhance the stochastic microstructure modeling of anodes presented in~\cite{feinauer.2015b} and~\cite{westhoff.2017} to the modeling of cathodes. In our case, the cathode material comes from pristine and cyclically aged Li-ion energy cells of the same type, where the material exhibits several (structural) differences compared to anodes, see Figure~\ref{fig:compAnodesCathode} for a visual impression. The main differences concern the volume fraction and the connectivity relations of the particle phase, and, very conspicuous, the more spherical-shaped cathode particles compared to anode particles.

Our enhanced model uses all basic concepts (i.e., random tessellations, connectivity graphs and the spherical harmonics series expansion of Gaussian random fields) introduced in~\cite{feinauer.2015b} and~\cite{westhoff.2017}, but each step of the stochastic model is adapted to meet the requirements of cathodes. By an appropriate alteration of the model parameters we are able to use this enhanced modeling approach not only for a pristine but also for two kinds of aged cathodes.

The paper has the following outline. In Section~\ref{sec:mat_data} we describe the tomographic image data of the cathode material and how the data is preprocessed. Then in Section~\ref{sec:model} the enhanced stochastic model is presented and it is shown how the model can be calibrated to the tomographic data. Section~\ref{sec:valid} contains an extensive model validation, where we compare tomographic and simulated data by means of several (image) characteristics. This is followed in Section~\ref{sec:discussion} by a discussion of differences in the morphologies of the pristine and the two aged cathodes. Finally, Section~\ref{sec:sum_out} gives a summary of the present work as well as a discussion of possible future prospects and tasks.

\section{Material and data}\label{sec:mat_data}

Tomographic image data of 3D microstructures as fundamental input for stochastic modeling was extracted from cathodes of an automotive plugin-hybrid energy cell. Thereby, besides pristine cathode material, two different aging scenarios of the same type of an energy cell cathode (NMC-cathode, Ni-rich, nominal capacity about 36.5Ah) were considered and for each scenario the materials' microstructure was measured by synchrotron tomography.

\subsection{Description of material and sample preparation}\label{subsec:material_prepara}

To go further into detail, we consider the cathode material for the following three scenarios: \textsc{P} - pristine plugin-hybrid energy cell (capacity 36.5Ah); \textsc{A1} - hybrid-electric-vehicle-profile aged cell (charge\hyp{}discharge\hyp{}balanced, capacity about 32Ah $\sim$ 90\% SOH); \textsc{A2} - electric-vehicle-profile aged cell (discharge\hyp{}dominated, capacity about 27.5Ah $\sim$ 80\% SOH). %; \textsc{A3} - 7\%-SOC-hub cyclization during high SOC-level (capacity about 27.5Ah $\sim$ 80\% SOH).
Note that the type of the energy cell cathode is always the same and the aged cells (\textsc{A1} and \textsc{A2}) were exposed to a steady cyclization over a period of about 6 months.
From each cell a layer (snippet) of cathode material was extracted and then all layers were stacked resulting in a multilayer sample stack of all scenarios, where the single layers were fixed and separated by a double-sided adhesive tape (type FL9628FL, manufacturer 3M). Finally, a smaller and rectangular sample was sliced from this larger multilayer sample stack to gain a suitable specimen for imaging. For details regarding the extraction and preparation of such a multilayer sample stack we refer to~\cite{mitsch.2014}.

\subsection{Data description and preprocessing}\label{subsec:data_prepro}

Using synchrotron tomography for imaging, it is possible to create 3D data sets of the extracted microstructures of the previously prepared specimen. This imaging was performed at the synchrotron X-ray facility BAMLine at BESSY in the Helmholtz\hyp{}Zentrum Berlin f\"ur Materialien und Energie (HZB). More detailed information on the procedure how the tomographic 3D data sets were created from the specimen, i.e., how the tomographic setup, measurement and finally the 3D data reconstruction were carried out, can also be found in~\cite{mitsch.2014}. This imaging procedure led to a three-dimensional grayscale image of the multilayer sample stack (16-bit, 3253$\times$2911$\times$2663 voxels) with a resolution of about $\unit[0.438]{\mu m}$ per voxel.

To make use of the extracted microstructures for later model calibration, some preprocessing of the tomographic 3D image data is necessary. First, for reasons of memory efficiency and computation time the 16-bit grayscale image was converted into an 8-bit grayscale image. Further, the single layers of each scenario were cut out from the imaged multilayer sample stack. Since, usually, the single layers are not planar but skewed and curved in the stack, we first applied a rotation and then a straightening using polynomial regression to obtain image data sets of almost planar sample layers for each scenario. Finally, we took a cutout from each sample layer of size 1000$\times$1000 voxels in horizontal direction and either $z_{\textsc{P}}=80$, $z_{\textsc{A1}}=105$ or $z_{\textsc{A2}}=100$ voxels in vertical direction. Due to damaged parts or artifacts in the imaged layers, especially close to the edges of a sample layer, the sizes of cutouts are restricted but still sufficiently large for model calibration. Therefore, we picked out every cutout from a preferably homogeneous position in the respective sample layer.

\begin{figure}[!ht]
	\centering
		\includegraphics[scale=0.5]{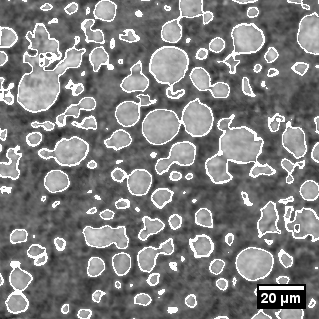}
	\caption{2D cutout of grayscale image (scenario \textsc{P}), where the boundaries of the particle phase after binarization are overlayed in white.}
	\label{fig:solidClusterValid}
\end{figure}

The good contrast in the grayscale images makes it possible to easily binarize the data, which means that each voxel is assigned via a global threshold value $t$ either to the particle phase (foreground, value 255) or to the pore phase (background, value 0). Note that the additives, like binder and carbon black, are considered as belonging to pore phase, because we cannot distinguish them in the tomographic images. Before global thresholding, the noise in the image data was reduced by applying a 3D median filter with a sphere of radius 3 as structuring element, see, e.g.,~\cite{burger.2008}. Then, we used the "Default" method for automatic thresholding implemented in the image processing program \textit{ImageJ} (Menu: Image $\triangleright$ Adjust $\triangleright$ Auto Threshold) to get an idea of suitable threshold values. This "Default" method is an iterative procedure based on the isodata algorithm, see~\cite{ridler.1978}. The global threshold values which were finally used to binarize the image data are listed in Table~\ref{tab:thresh} and are between 2 to 5 units lower than the values suggested by the automatic "Default" method. We adjusted the suggested values because they seemed to be a bit too high. By visual inspection the binarization for the lower values hit the bounds of the visible particle phase in the grayscale images better, see Figure~\ref{fig:solidClusterValid}.
\begin{table}[!t]
	\centering
	\begin{tabular}{r|c|c|c}
		\hline
		& \multicolumn{3}{c}{scenario}\\
		& \textsc{P} & \textsc{A1} & \textsc{A2}\\
		\hline
	%\begin{tabular}{rccc}%c}
		%\hline
		%image data of scenario & \textsc{P} & \textsc{A1} & \textsc{A2}\\% & \textsc{A3}\\
		%\hline
		global threshold value $t$ & 120 & 127 & 138\\% & 139\\
		\hline
	\end{tabular}
	\caption{Global threshold values applied to the filtered grayscale images for each scenario.}
	\label{tab:thresh}
\end{table}
Next, as we do not want the particles to have holes, all small isolated pore clusters in the particle phase are set to the particle phase using the same approach as in~\cite{feinauer.2015b} and~\cite{westhoff.2017}. Since we could not detect isolated pore clusters larger than 5000 voxels, we set all those pore clusters smaller than this value to the particle phase. As a final correction step regarding the binarization of the image data, we removed a few small artifacts from the particle phase, i.e., all by the Hoshen\hyp{}Kopelman clustering algorithm (see~\cite{hoshen.1976}) detected particle clusters smaller than 100 voxels were set to the pore phase. Such small particle clusters usually result from errors in sample preparation and imaging or are impurities or negligible small particle fragments.

Finally, as the stochastic model considered in Section~\ref{sec:model} is a so-called particle-based model and therefore needs the information on individual particles of the cathode microstructure, it is necessary to segment the binarized image data. For this purpose, we performed exactly the same marker-based watershed algorithm as in~\cite{feinauer.2015b} and~\cite{westhoff.2017} to detect and then label each particle by a unique integer.
\begin{figure}[!ht]
	\centering
	%\begin{subfigure}[c]{0.49\textwidth}
		\includegraphics[width=\textwidth]{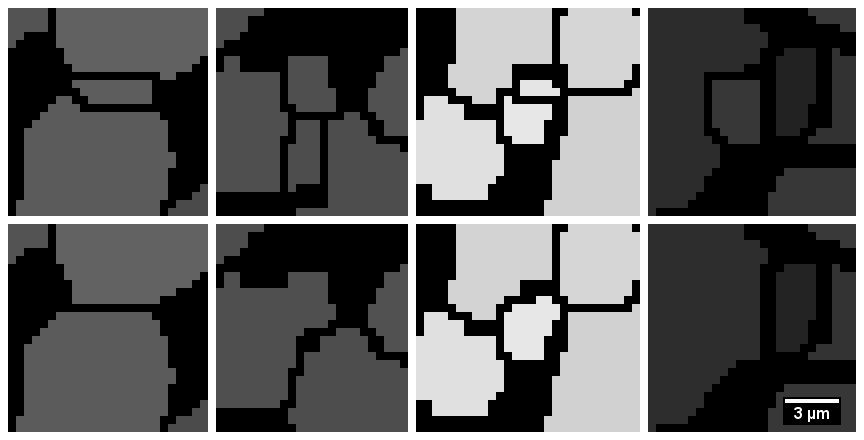}
		%\caption{before post-segmentation}
		%\label{subfig:postseg_before}
	%\end{subfigure}
	%\begin{subfigure}[c]{0.49\textwidth}
		%\includegraphics[width=\textwidth]{images/postseg_after.png}
		%\caption{after post-segmentation}
		%\label{subfig:postseg_after}
	%\end{subfigure}
	\caption{Examples of 2D cutouts for particles before (first row) and after post-segmentation (second row).}
	\label{fig:postseg}
\end{figure}
To improve the segmentation results, we additionally performed a post-segmentation step, since there were still some oversegmented areas, especially regarding small fragments between particle pairs, see first row of Figure~\ref{fig:postseg}. For the post-segmentation step we iterate over all previously segmented particles, dilate the currently considered particle by a ball of some radius $r_{post}\geq 0$, check then if the dilated particle completely covers adjacent particles and remove the markers of these adjacent particles from the marker set. In the case that two dilated particles completely cover each other, then actually just the marker of the (originally) smaller particle is removed. Finally, we use this subset of markers to restart the above-mentioned watershed algorithm again. The radius $r_{post}=7$ worked very well for our data as one can see in Figure~\ref{fig:postseg}. A final segmentation result is shown in Figure~\ref{subfig:pov}.

Now, as all preprocessing is done and all essential information gathered, we can turn to the description of the stochastic microstructure model and its calibration to tomographic image data.

\section{Stochastic 3D microstructure modeling}\label{sec:model}

The stochastic model for the microstructure of cathodes is based on modeling ideas which have recently been introduced for anodes in~\cite{feinauer.2015b} and~\cite{westhoff.2017}. However, to account for the structural differences between anodes and cathodes, some steps of the modeling procedure had to be modified. For this purpose, tools from stochastic geometry are used, in particular various kinds of random (marked) point processes and random tessellations, see, e.g.,~\cite{chiu.2013}.

\begin{figure}[!ht]
	\centering
	\begin{subfigure}[c]{0.325\textwidth}
		\includegraphics[width=\textwidth]{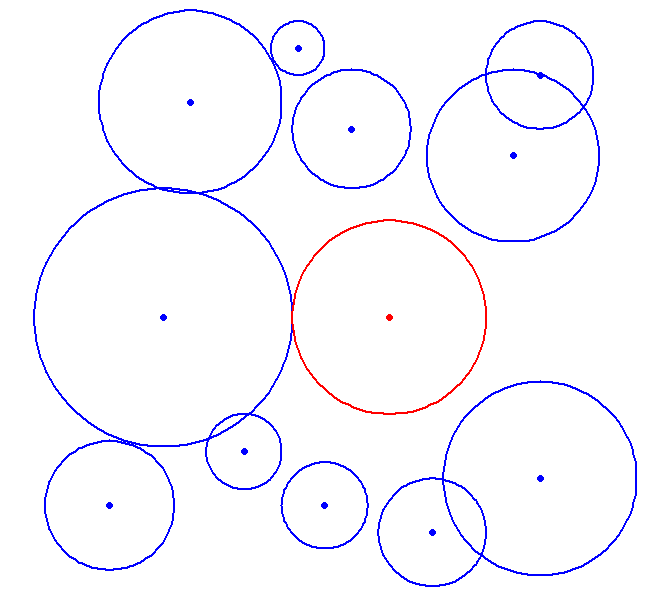}
		\caption{}
		\label{subfig:pointPattern}
	\end{subfigure}
	\begin{subfigure}[c]{0.325\textwidth}
		\includegraphics[width=\textwidth]{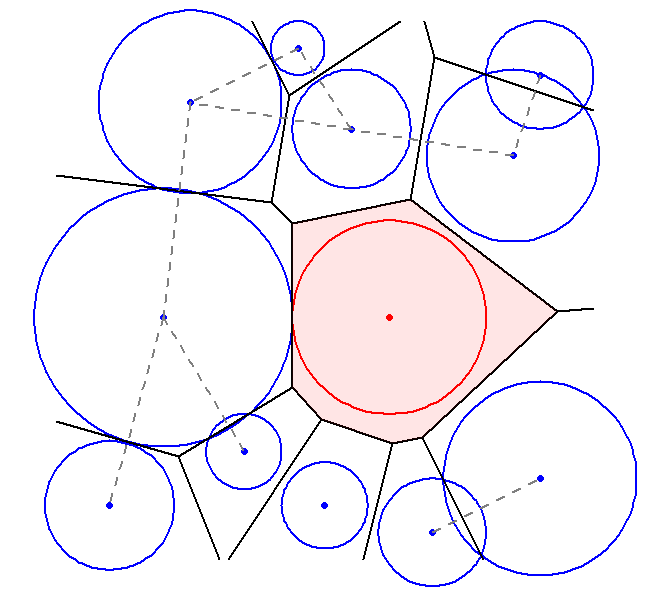}
		\caption{}
		\label{subfig:tessAndGraph}
	\end{subfigure}
	\begin{subfigure}[c]{0.325\textwidth}
		\includegraphics[width=\textwidth]{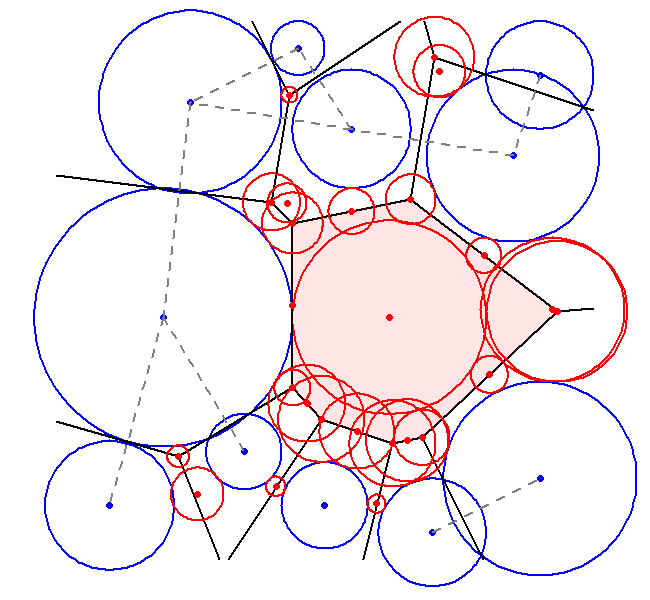}
		\caption{}
		\label{subfig:poreGenerators}
	\end{subfigure}
	\begin{subfigure}[c]{0.325\textwidth}
		\includegraphics[width=\textwidth]{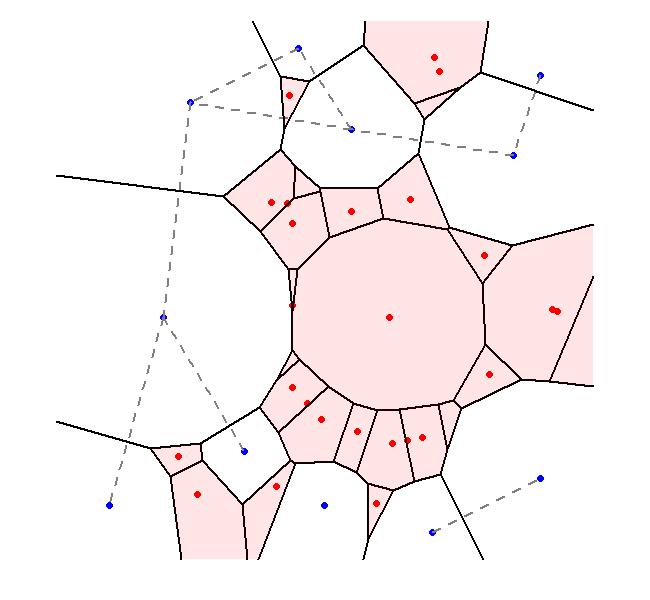}
		\caption{}
		\label{subfig:tessFinal}
	\end{subfigure}
	\begin{subfigure}[c]{0.325\textwidth}
		\includegraphics[width=\textwidth]{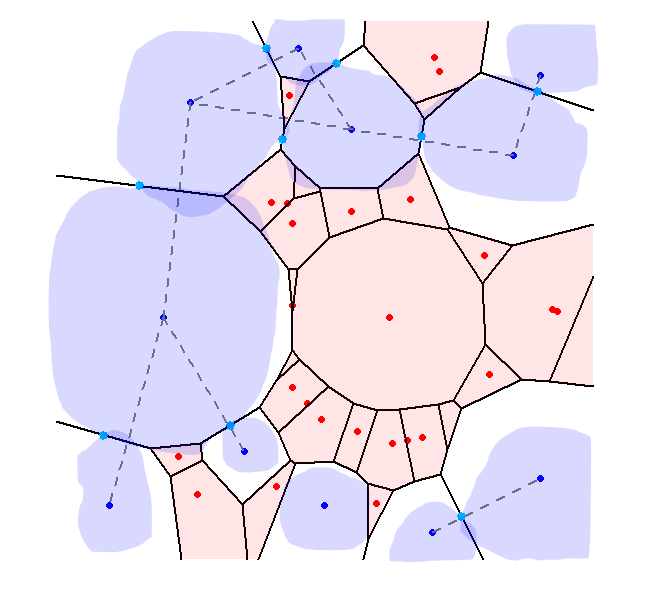}
		\caption{}
		\label{subfig:tessFinalAndParticles}
	\end{subfigure}
	\begin{subfigure}[c]{0.325\textwidth}
		\includegraphics[width=\textwidth]{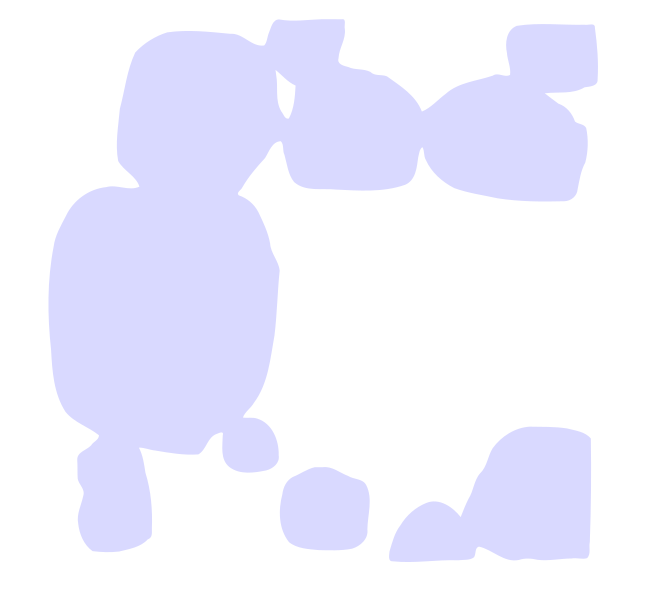}
		\caption{}
		\label{subfig:particles}
	\end{subfigure}
	\caption{Overview of the modeling steps as 2D sketch. (a) Two random marked point patterns are realized, where the blue dots and circles induce particles and the red ones induce large pores; (b) A connectivity graph (dashed gray lines) based on the random marked point patterns and the corresponding Laguerre tessellation (black lines) is simulated, where the red shaded Laguerre polytope indicates an (empty) pore polytope (i.e., no particle is placed into); (c) Additional marked points (further red dots and circles) are determined that induce further pore polytopes; (d) Final arrangement of particle polytopes (i.e., a particle is placed into) and pore polytopes (red shaded) is computed, where the initial connectivity is still retained; (e) Particles fulfilling contact conditions (light blue dots) are created in the corresponding polytopes using Gaussian random fields on the sphere; (f) Only the particles are kept and morphological smoothing operations lead to the final microstructure.}
	\label{fig:modelSketch}
\end{figure}

\subsection{Basic modeling ideas}\label{subsec:basicIdeas}

The construction of the model starts with the introduction of two random marked point patterns, see Figure~\ref{subfig:pointPattern}. Due to very low volume fraction and locally occurring large pores in the considered cathodes, one random marked point pattern particularly models such large pores, where the marks indicate the approximate pore radii and the points their locations. The other random marked point pattern determines the approximate particle locations and sizes. Next, these two sets of marked points are used to decompose the region of interest (in general a 3-dimensional cuboid) into a set of convex polytopes by means of a Laguerre tessellation (black lines in Figure~\ref{subfig:tessAndGraph}). Thereby, convex polytopes remain empty which are induced by marked points modeling large pores, i.e., later on no particles will be placed in these (pore) polytopes. The red shaded polytope in Figure~\ref{subfig:tessAndGraph} indicates such a pore polytope. Based on such a tessellation, we model connectivity between particles depending on the surface area of a Laguerre facet between two corresponding polytopes in which particles will be placed. %To keep the stochastic model not too complex and as this dependency is enough, we do not relate the connectivity of particles to further dependencies like, e.g., the distance between two points of the first pattern which induce the polytopes for particles.
A distance ratio of the two marked points which had induced these neighboring polytopes is the second quantity on which connectivity between particles depends. The resulting connectivity graph, see dashed gray lines in Figure~\ref{subfig:tessAndGraph}, is not necessarily fully connected which we do not require for particle systems of cathodes contrary to anodes, where we assumed full connectivity of all particles. At this modeling stage, we are confronted with the problem that the polytopes into which particles will be placed, see transparent polytopes in Figure~\ref{subfig:tessAndGraph}, are too large and badly formed to gain reasonably shaped particles, especially nearly spherical-shaped particles. Therefore, we introduce a third set of marked points inducing further pore polytopes which remain empty and simultaneously reduce the volume fraction of polytopes for particles. For this purpose, we look for candidates of additional marked points. For each candidate, to be added to the generators of the Laguerre tessellation (red and blue circles in Figure~\ref{subfig:pointPattern}), it is checked whether it induces a polytope which covers some predefined restriction points (see Section~\ref{subsubsec:porePoly} for details). If not, then we add this candidate to the third set of marked points generating polytopes which remain empty (see red circles in Figure~\ref{subfig:poreGenerators}). Otherwise, the candidate point is marked with the half of its previous mark and is again checked as described above. If this (modified) candidate still induces a polytope which covers some of the restriction points, then the candidate point is marked with 0. If it is still not possible to add it, then this candidate point is rejected. This procedure is performed for all candidate points. Finally, we obtain a tessellation with suitably sized and reasonably shaped polytopes for particles, see transparent polytopes in Figure~\ref{subfig:tessFinal}. In these polytopes particles are placed fulfilling contact conditions given by the connectivity graph (see Figure~\ref{subfig:tessFinalAndParticles}). The particles themselves are realizations of Gaussian random fields on the sphere. The final result shown in Figure~\ref{subfig:particles} exhibits the desired large pores and the typically nearly spherical-shaped particles.

In the following, all modeling steps summarized above are discussed in detail.

\subsection{Modeling the approximate configuration of large pores and particles}\label{subsec:pointModel}

\subsubsection{Initial systems of spheres}\label{subsubsec:initSphereSystems}

\begin{figure}[!t]
	\centering
	\begin{subfigure}[t]{0.49\textwidth}
		\includegraphics[width=\textwidth]{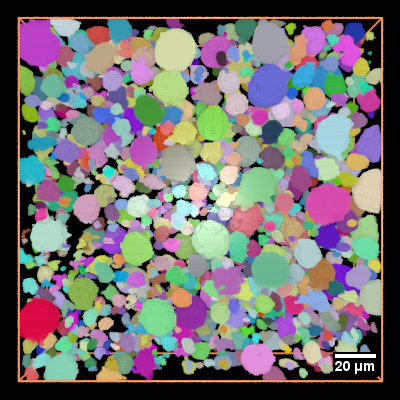}
		\caption{Segmented structure}
		\label{subfig:pov}
	\end{subfigure}
	\begin{subfigure}[t]{0.49\textwidth}
		\includegraphics[width=\textwidth]{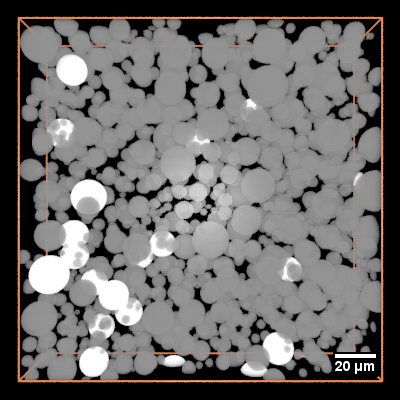}
		\caption{Segmented structure as spheres}
		\label{subfig:balls}
	\end{subfigure}
	\caption{3D renderings of structure cutouts. (a) Particles of the pristine cathode (scenario \textsc{P}) after segmentation; (b) Particles represented as gray spheres with volume-equivalent radii around centers of mass and pores of radius greater or equal $t_p=15.0$ voxels depicted as white spheres with corresponding minimum pore radii.}
	\label{fig:pov_balls}
\end{figure}

\begin{figure}[!ht]
	\centering
	\begin{subfigure}[c]{0.49\textwidth}
		\includegraphics[width=\textwidth]{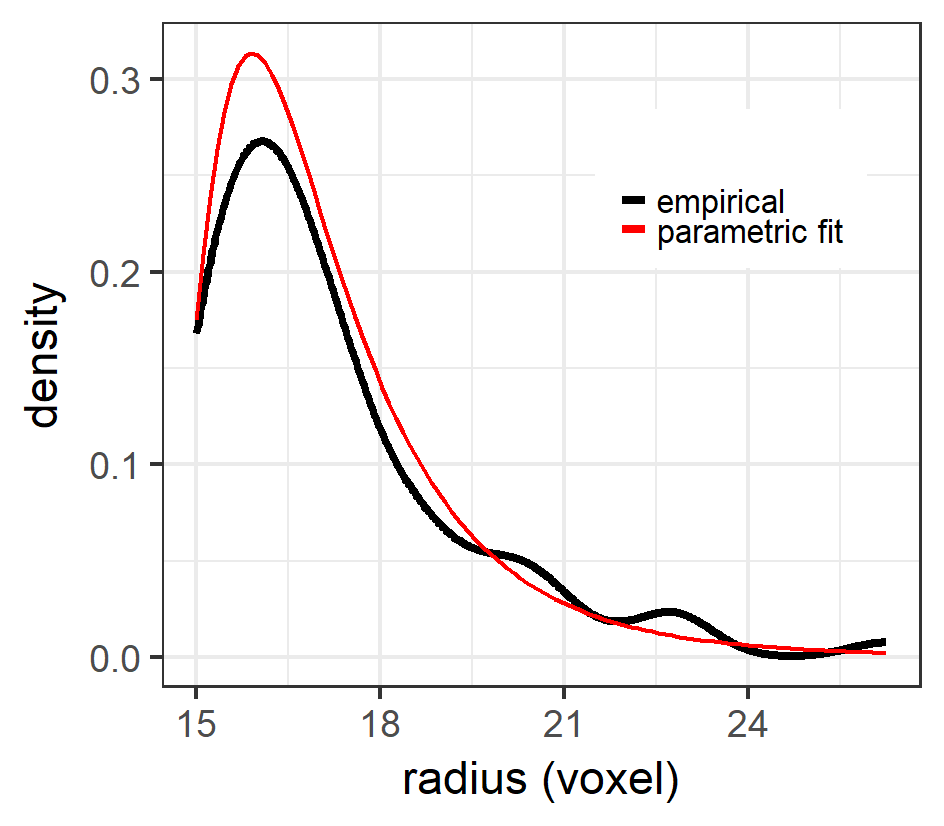}
		\caption{Minimum pore radii of large pores}
		\label{subfig:poreRadiiP}
	\end{subfigure}
	\begin{subfigure}[c]{0.49\textwidth}
		\includegraphics[width=\textwidth]{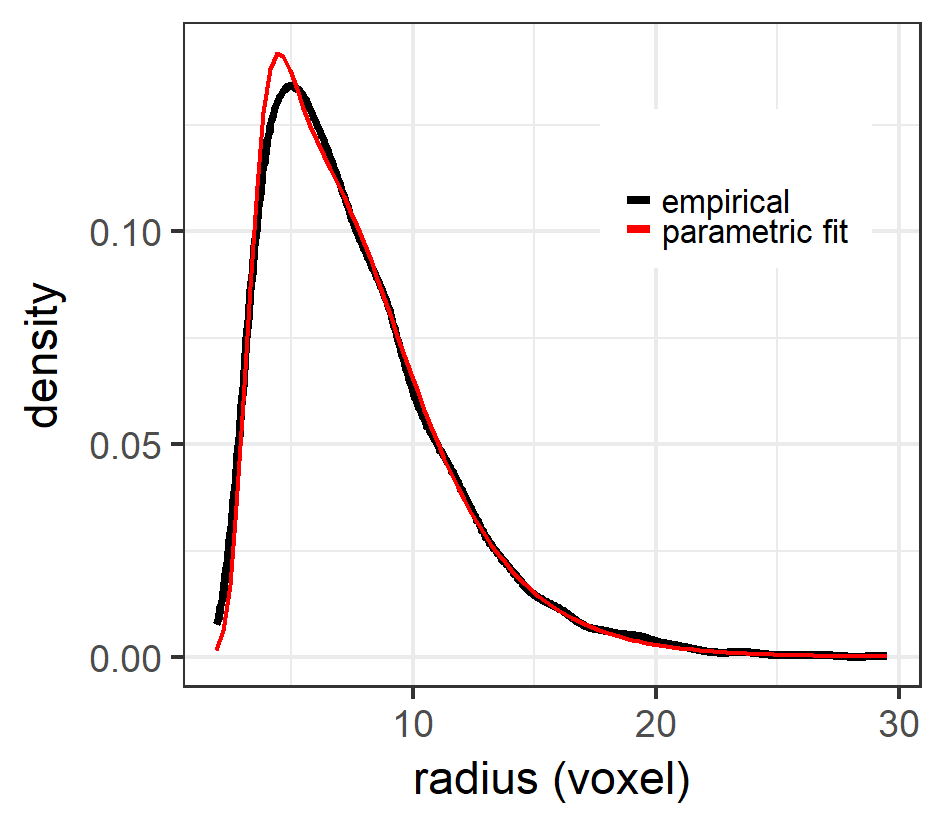}
		\caption{Volume-equivalent radii of particles}
		\label{subfig:particleRadiiP}
	\end{subfigure}
	\caption{Empirical distribution of minimum pore radii greater or equal $t_p=15.0$ voxels and the parametric probability distribution fit (left); empirical distribution of volume-equivalent particle radii and the parametric probability distribution fit (right). Both empirical distributions are extracted from the pristine cathode (scenario \textsc{P}).}
	\label{fig:particle_pore_radiiP}
\end{figure}

As mentioned at the beginning of Section~\ref{subsec:basicIdeas}, due to low volume fraction and locally large pores in the considered cathodes, we introduce a random marked point pattern which particularly models the large pores. For that reason, we extract from the (binarized) tomographic image data just pores with a (minimum) pore radius greater or equal $t_p>0$. We call $t_p$ a pore threshold. Figure~\ref{subfig:balls} depicts such large pores as a system of spheres (white), where the spheres are located at pores with corresponding minimum pore radii. The empirical distribution function of minimum pore radii greater or equal $t_p$ extracted from binarized tomographic image data of pristine cathode material is shown in Figure~\ref{subfig:poreRadiiP}, black curve. It turns out that this empirical distribution can be approximated by a truncated (and then shifted) log-mixed-normal distribution, see red curve in Figure~\ref{subfig:poreRadiiP}. Note that, if we have a mixed-normal random variable $Z^\prime$ with mean values $\mu_1^\prime, \mu_2^\prime\in\R$, standard deviations $\sigma_1^\prime, \sigma_2^\prime>0$ and probability mix parameter $0\leq\alpha^\prime\leq 1$, then $R^\prime=\exp(Z^\prime)$ is said to be log-mixed-normally distributed. Additionally, when restricting the values of $R^\prime$ to an interval $[l^\prime, u^\prime]$, then $R^\prime$ is said to be truncated log-mixed-normal with truncation bounds $l^\prime>0$ and $u^\prime>l^\prime$. Finally, $R^\prime_{s^\prime}=R^\prime+s^\prime$ with $s^\prime\in\R$ is a shifted version of $R^\prime$.

Before we continue with the random point pattern model for large pores, we explain how pores (i.e., their locations) with minimum pore radii greater or equal $t_p$ are extracted from tomographic data. First, we consider the Euclidean distance transformation $E\colon\R^3\to\R_+$ on the pore phase $\Xi\subset\R^3$ of a binarized tomographic data set, where we assume that $\Xi$ is not the empty set. That is, the value $E(x)$ for $x\in\Xi$ gives the minimum Euclidean distance to the particle phase $\Xi^c=\R^3\setminus\Xi$, where $E(x)=0$ for $x\in\Xi^c$. Then, the set of locations \[X^*=\{x^*\in\Xi:E(x^*)\geq E(x) \text{ for some } \varepsilon>0 \text{ and all } x\in B(x^*,\varepsilon)\}\] of local maxima of $E$ are detected, where $B(y,e)$ denotes a sphere around center $y$ with radius $e$.
%If a local maximum is reached in more points which are in "direct neighborhood", then we take the centroid of these points as location for this local maximum. Note that if the centroid should not belong to these points itself, then the closest of these point to the centroid is taken.
Note that the concept of local maxima (or vice versa of local minima if one would consider the negative Euclidean distance transform $-E$) for voxel-based image data has been considered, e.g., in~\cite{spettl.2015}. Next, as we just want to model large pores, only local maxima $\{E(x^*), x^*\in X^*\}$ greater or equal $t_p>0$ are considered, which we briefly denote by $\{E(x^*)>t_p\}$. Finally, $\{E(x^*)>t_p\}$ is thinned out since the detection of local maxima is quite sensitive to "small artifacts" in binarized tomographic image data. The rule for thinning is as follows:
\begin{enumerate}[nolistsep, leftmargin=20pt]
\item[(a)] We iterate over all local maxima $\{E(x^*)>t_p\}$;
\item[(b)] For the current local maximum we consider all other local maxima around whose distance to the current maximum is at most $2t_p$;
\item[(c)] If for one of these neighboring local maxima it holds that this local maximum is greater or equal to the current local maximum, then (the location of) the current local maximum will not be added to the set $X^{**}=\{x^{**}\in X^*: E(x^{**})>t_p\}$.
\end{enumerate}
Recall that the values $\{E(x^{**})\}$ exactly correspond to the minimum pore radii mentioned above.
%The result is a system of spheres representing the large pores, where each sphere is located at a local maximum $x^{**}$ with radius $E(x^{**})$.

Because of the applied rule for thinning, the point pattern of pore locations $X^{**}$ has a hard-core distance of $2t_p$, i.e., $X^{**}\cap B(x^{**},2t_p)=x^{**}$ for all $x^{**}\in X^{**}$. Therefore, it is reasonable to model the points in $X^{**}$ via a Mat\'{e}rn hard-core point process $\{\tilde{S}_n,n\in\N\}$ with some intensity $\tilde{\lambda}>0$ and hard-core radius $\tilde{r}_h>0$, see, e.g.,~\cite{chiu.2013}. Furthermore, we mark the points of $\{\tilde{S}_n\}$ with sequence $\{\tilde{R}_n,n\in\N\}$ of independent random variables, which are also independent of the sequence $\{\tilde{S}_n\}$. Each (minimum pore) radius $\tilde{R}_n$ is drawn from the above introduced truncated and shifted log-mixed-normal distribution with some parameters $\tilde{\mu}_1,\tilde{\mu}_2,\tilde{\sigma}_1,\tilde{\sigma}_2,\tilde{\alpha},\tilde{l}, \tilde{u}$ and $\tilde{s}$, where now $0\leq \tilde{s}<t_p$.

The sequence of pairs $\{(\tilde{S}_n,\tilde{R}_n)\}$ constitutes the random marked point pattern which models the large pores. Its representation as a system of spheres $\{B(\tilde{S}_n,\tilde{R}_n)\}$ has also influence on the collective rearrangement algorithm described in Section~\ref{subsubsec:collRearrang}.

The next step in stochastic 3D microstructure modeling of cathodes is to find a model for marked point patterns which describes the (initial) approximate particle locations and sizes. For this purpose, we use a random system of slightly overlapping spheres in exactly the same manner as already described in~\cite{westhoff.2017}.
%The sphere midpoints and radii shall correspond to approximate locations and volume-equivalent radii of the later realized particles.
Again, it is necessary that we represent each segmented particle in the tomographic 3D image data from Section~\ref{subsec:data_prepro}, see Figure~\ref{subfig:pov}, by a volume-equivalent sphere. That is, each particle is transformed into a sphere with volume-equivalent radius $\sqrt[3]{3V/4\pi}$ around its center of mass, see gray spheres in Figure~\ref{subfig:balls}, where $V$ is the volume of the corresponding particle. To model such an extracted system of slightly overlapping (volume-equivalent) spheres, we begin with a random marked point pattern which can be interpreted as an initial system of (volume-equivalent) spheres. This initial system is later used as input for a collective rearrangement algorithm (also known as force-biased algorithm, see, e.g.,~\cite{moscinski.1989}). These algorithms have proved to be successful for the modeling of sphere systems with nearly vanishing overlaps which is the case for our extracted system of spheres.

As a model for the random marked point pattern we apply a Mat\'{e}rn soft-core process~\cite{stoyan.1987} with some intensity $\lambda>0$ and soft-core radius (i.e. random hard-core radius) $R>0$ following the truncated log-mixed-normal distribution with some parameters $\mu_1,\mu_2,\sigma_1,\sigma_2,\alpha,l$ and $u$ introduced above. This point process model has turned out to be appropriate to generate initial systems of (volume-equivalent) spheres for the subsequent collective rearrangement algorithm. Thus, for the initial approximate particle locations and sizes we utilize a measurable indexing $\{(S_n,R_n),n\in\N\}$ of the Mat\'{e}rn soft-core process. The independent marks $\{R_n\}$, which are also independent of the points $\{S_n\}$, mimic the volume-equivalent particle radii extracted from segmented tomographic image data. Figure~\ref{subfig:particleRadiiP} shows that the parametric distribution of $R$ (red curve) fits the empirical distribution of volume-equivalent particle radii (black curve) quite well. Altogether, the result is a random marked point pattern $\{(S_n,R_n)\}$ or in other words an initial systems of (volume-equivalent) spheres $\{B(S_n,R_n),n\in\N\}$ for the following collective rearrangement algorithm.

%Further details on random (marked) point processes can be found, e.g., in~\cite{chiu.2013}.

\subsubsection{Collective rearrangement algorithm}\label{subsubsec:collRearrang}

In practice, the realization of samples of the random marked point patterns $\{(S_n,R_n)\}$ and $\{(\tilde{S}_n,\tilde{R}_n)\}$ introduced in Section~\ref{subsubsec:initSphereSystems} is restricted to a bounded sampling window $W\subset\R^3$, where we apply periodic boundary conditions to maintain the properties of the point processes. For the following collective rearrangement algorithm, which is performed on $W$, the representation of the point patterns as systems of spheres $\{B(S_n,R_n)\}$ and $\{B(\tilde{S}_n,\tilde{R}_n)\}$ serves as initial input. Note that the spheres $\{B(\tilde{S}_n,\tilde{R}_n)\}$ which will induce large pores are not rearranged by the algorithm and that the algorithm itself is a modification of well-known rearrangement algorithms, where we refer, e.g., to~\cite{hirsch.2015, moscinski.1989} for details. It works as follows.
%The simulation of the above two random marked point patterns $\{(S_n,R_n)\}$ and $\{(\tilde{S}_n,\tilde{R}_n)\}$ is executed on a dilated sampling window $W^\prime\supset W$ regarding the elimination of edge effects. Further, their representation as spheres $\{B(S_n,R_n)\}$ and $\{B(\tilde{S}_n,\tilde{R}_n)\}$ is the initial configuration of particles and large pore for the following collective rearrangement algorithm.
\begin{enumerate}[leftmargin=*]
\setlength{\itemsep}{0pt}
	\item\label{1} For each sphere $B(S_n,R_n)$ compute the random shift vector \[F_n=\sum\limits_{S_m\neq S_n}{\frac{O_m}{2}\frac{(S_n-S_m)}{\|S_n-S_m\|}\ind_{\{O_m>0\}}}+\sum\limits_{\tilde{S}_m}{\tilde{O}_m\frac{(S_n-\tilde{S}_m)}{\|S_n-\tilde{S}_m\|}\ind_{\{\tilde{O}_m>0\}}},\] where $O_m=R_n+R_m-\|S_n-S_m\|$ and $\tilde{O}_m=R_n+\tilde{R}_m-\|S_n-\tilde{S}_m\|$ are random lengths of overlaps with spheres $B(S_m,R_m)$ and $B(\tilde{S}_m,\tilde{R}_m)$, respectively. In other words, for all intersections with the sphere $B(S_n,R_n)$, we sum up all unit vectors from the midpoint of an intersecting sphere to $S_n$ scaled to the length of half overlap $O_m$ or total overlap $\tilde{O}_m$. Here, $\ind_{\{O>0\}}$ is the indicator function of a random overlap length $O$ being greater than 0 and $\|S-S^\prime\|$ denotes the Euclidean distance between some points $S$ and $S^\prime$.
	\item For each sphere $B(S_n,R_n)$ compute the random mean overlap $\bar{O}_n$ with all other spheres $\{B(S_m,R_m), m\neq n\}$ which intersect with it, i.e.,
	\[\bar{O}_n=\frac{\sum_{m\neq n}{O_m\ind_{\{O_m>0\}}}}{\#\{O_m>0, m\neq n\}},\] 
	%and the random mean overlap $\bar{\tilde{O}}_n$ with all spheres $\{B(\tilde{S}_m,\tilde{R}_m)\}$ analogously, i.e., \[\bar{\tilde{O}}_n=\frac{\sum_{m}{\tilde{O}_{nm}\ind_{\{\tilde{O}_{nm}>0\}}}}{\#\{\tilde{O}_{nm}>0\}},\]
	where $\#$ denotes the cardinality of a set, i.e., the number of elements in a set.
	\item If the mean of all random mean overlaps $\bar{O}_n$ is greater than a given threshold $t_O$% and if the mean of all random mean overlaps $\bar{\tilde{O}}_n$ is greater than a given threshold $t_{\tilde{O}}$
, then shift each sphere $B(S_n,R_n)$ by its random vector $F_n$ and restart with step~\ref{1}. Otherwise terminate the algorithm.
\end{enumerate}
Note that if a midpoint $S_n$ of the sphere $B(S_n,R_n)$ is shifted outside the sampling window $W$, then it will be displaced to enter $W$ on the opposite side to maintain the condition of periodic boundaries. The length of each shift vector $F_n$ is chosen in such a way that the initial systems of spheres are successively transformed into an overall sphere system which finally consists of nearly non-overlapping spheres. That is, the achieved mean overlap among spheres $B(S_n,R_n)$, $n\in\N$ is approximately $t_O$ and the achieved mean overlap between spheres $B(S_n,R_n)$, $n\in\N$ and $B(\tilde{S}_n,\tilde{R}_n)$, $n\in\N$ is almost zero.% close to $t_{\tilde{O}}$.

The outcome of the collective rearrangement algorithm is a marked point pattern which constitutes the approximate configuration of large pores and particles. We denote it by $\mathcal{S}\cup\tilde{\mathcal{S}}$ in the subsequent modeling steps, where $\mathcal{S}=\{(S_n,R_n)\}$ and $\tilde{\mathcal{S}}=\{(\tilde{S}_n,\tilde{R}_n)\}$.

\subsubsection{Calibration of point pattern models}\label{subsubsec:caliPointPatternModels}

To calibrate the point pattern models to the tomographic image data of scenarios \textsc{P}, \textsc{A1} and \textsc{A2}, respectively, it is necessary to suitably determine the values of all model parameters, see Table~\ref{tab:pointParam}.
\begin{table}[!t]
	\centering
	%\begin{tabular}{cc*{11}{|c}}
	\begin{tabular}{r|c|c|c}
		\hline
		%& & $t_p$ & $\tilde{\lambda}$ & $\tilde{r}_h$ & $\tilde{\mu}_1$ & $\tilde{\mu}_2$ & $\tilde{\sigma}_1$ & $\tilde{\sigma}_2$ & $\tilde{\alpha}$ & $\tilde{l}$ & $\tilde{u}$ & $\tilde{s}$\\% & $t_{\tilde{O}}$\\
		%\hline
		%\multirow{3}{*}{\rotatebox[origin=c]{90}{scenario}} & \textsc{P} & 15 & $1.3\cdot 10^{-6}$ & $2\cdot t_p$ & 0.99 & - & 0.59 & - & 1 & 1 & 12.25 & 14\\% & 0.014\\
		%& \textsc{A1} & 15 & $1.09\cdot 10^{-6}$ & $2\cdot t_p$ & 1.93 & 0.8 & 0.12 & 0.46 & 0.16 & 1 & 8.41 & 14\\% & 0.014\\
		%& \textsc{A2} & 13 & $1.74\cdot 10^{-6}$ & $2\cdot t_p$ & 0.82 & - & 0.52 & - & 1 & 1 & 14.72 & 12\\% & 0.024\\
		& \multicolumn{3}{c}{scenario}\\
		& \textsc{P} & \textsc{A1} & \textsc{A2}\\
		\hline
		$t_p$ & 15.0 & 15.0 & 13.0\\
		$\tilde{\lambda}$ & $1.30\cdot 10^{-6}$ & $1.09\cdot 10^{-6}$ & $1.74\cdot 10^{-6}$\\
		$\tilde{r}_h$ & $2.0\cdot t_p$ & $2.0\cdot t_p$ & $2.0\cdot t_p$\\
		$\tilde{\mu}_1$ & 0.99 & 1.93 & 0.82\\
		$\tilde{\mu}_2$ & - & 0.80 & -\\
		$\tilde{\sigma}_1$ & 0.59 & 0.12 & 0.52\\
		$\tilde{\sigma}_2$ & - & 0.46 & -\\
		$\tilde{\alpha}$ & 1.0 & 0.16 & 1.0\\
		$\tilde{l}$ & 1.0 & 1.0 & 1.0\\
		$\tilde{u}$ & 12.25 & 8.41 & 14.72\\
		$\tilde{s}$ & 14.0 & 14.0 & 12.0\\
		%$t_{\tilde{O}}$ & 0.014 & 0.014 & 0.024\\
		\hdashline
		$\lambda$ & $9.95\cdot 10^{-5}$ & $9.33\cdot 10^{-5}$ & $1.11\cdot 10^{-4}$\\
		$\mu_1$ & 1.44 & 2.13 & 1.93\\
		$\mu_2$ & 2.07 & 1.46 & -\\
		$\sigma_1$ & 0.24 & 0.41 & 0.49\\
		$\sigma_2$ & 0.41 & 0.26 & -\\
		$\alpha$ & 0.21 & 0.78 & 1.0\\
		$l$ & 1.91 & 2.03 & 1.86\\
		$u$ & 29.51 & 32.83 & 32.88\\
		$t_O$ & 0.657 & 0.770 & 0.634\\
		\hline
	\end{tabular}
	\caption{Parameter values of the point pattern models. The parameters $t_p, \tilde{r}_h, \tilde{l}, \tilde{u}, \tilde{s}, %t_{\tilde{O}}, 
	l, u$ and $t_O$ are given in voxel length. Empty fields (-) result from the fact that the probability mix parameter $\tilde{\alpha}$ or $\alpha$ is 1 and thus there is no second mean and standard deviation.}
	\label{tab:pointParam}
\end{table}
The intensities $\tilde{\lambda}$ and $\lambda$ are estimated from the extracted point patterns given by pores (with minimum pore radii greater or equal $t_p$) and particle centers of mass in the tomographic data sets. The pore threshold $t_p$ is approximately chosen as the radius of a sphere such that 10\% of the pore phase in a binarized tomographic data set can be covered with spheres of that radius (see~\cite{muench.2008} for details). The choice of the hard-core radius $\tilde{r}_h$ is a direct consequence of the above mentioned thinning of local maxima. All parameters of the log-mixed-normal distributions are estimated via the \textsf{R}-implemented \textit{mixtools} package~\cite{benaglia.2009} (based on expectation maximization), if we have two mixture components or via the \textsf{R}-implemented \textit{fitdistrplus} package~\cite{delignette.2015} (based on maximum likelihood estimation) provided that we only have a single mixture component. The limits $\tilde{l}$ and $\tilde{u}$ are set to $t_p-\tilde{s}$ and $r_{pmax}-\tilde{s}$, where $r_{pmax}$ is the maximum of all minimum pore radii $\{E(x^{**})\}$ observed in the corresponding tomographic data set. The shift parameter $\tilde{s}$ is given by $\tilde{s}=\lfloor r_{pmin}\rfloor-1$, where now $r_{pmin}$ is the minimum of all minimum pore radii. The expression $\lfloor v\rfloor$ means the greatest integer less than or equal to $v\in\R$. The parameters $l$ and $u$ are the minimum and maximum volume-equivalent particle radius observed in the corresponding segmented tomographic data set. The overlap threshold $t_O$ is chosen as the mean of all mean overlaps among particles represented as spheres with volume-equivalent radii, see e.g., white spheres in Figure~\ref{subfig:balls}.
%The second overlap threshold $t_{\tilde{O}}$ is calculated as the mean of all mean overlaps between particles again represented as spheres with volume-equivalent radii and large pores represented as spheres with minimal pore radii greater or equal $t_p$, see, e.g., white and gray spheres in Figure~\ref{subfig:balls}.

\subsection{Modeling the connectivity graph}\label{subsec:graphModel}

Recall that in Section~\ref{subsec:pointModel} we have determined the approximate configuration of particles and large pores. Since all considered microstructures consist of connected particle systems, the logical next step is to determine which particles are supposed to be connected. Thus, we make use of a random graph $G=(\mathcal{V},\mathcal{E})$ which shall describe the connectivity relations between the particles to be simulated later on, where $\mathcal{V}$ is a (random) set of vertices and $\mathcal{E}\subset\mathcal{V}\times\mathcal{V}$ are randomly placed edges (segments) between some pairs of the vertices. In our context, as vertices of the graph we use the marked point pattern considered in Section~\ref{subsec:pointModel}, which describes the approximate particle locations and sizes, i.e., $\mathcal{V}=\mathcal{S}$.
%Note that the marked points in $\tilde{\mathcal{S}}$ are no vertices of our graph since they model the large pores and, of course, will not belong to the connected system of particles.
The edges between some pairs of these marked points indicate where the corresponding particles are supposed to be connected.

\subsubsection{Particle connections based on a Laguerre tessellation}\label{subsubsec:particleConnections}

In the following, we briefly explain how we model the connectivity graph since the basic ideas are similar to those in~\cite{feinauer.2015b} and~\cite{westhoff.2017}, where especially in~\cite{westhoff.2017} further details can be found. We start from a Laguerre tessellation $\mathcal{T}_1$ on the sampling window $W$ which is induced by the marked point pattern $\mathcal{S}\cup\tilde{\mathcal{S}}$ from Section~\ref{subsec:pointModel}, i.e., this tessellation is a collection of convex polytopes $\mathcal{T}_1=\{P_n\}\cup\{\tilde{P}_n\}$, where the convex polytopes $P_n$ and $\tilde{P}_n$ correspond to the generator points $(S_n,R_n)\in\mathcal{S}$ and $(\tilde{S}_n,\tilde{R}_n)\in\tilde{\mathcal{S}}$, respectively. For further details on Laguerre tessellations we refer to~\cite{lautensack.2008}. Note that $\{P_n\}$ are particle polytopes, i.e., convex polytopes into which particles will be placed, and $\{\tilde{P}_n\}$ are pore polytopes, i.e., convex polytopes which remain empty. If $P_n$ is the particle polytope of generator $(S_n,R_n)\in\mathcal{S}$ and $P_m$ of $(S_m,R_m)\in\mathcal{S}$, then by $F_{nm}$ we denote the joint (two-dimensional) Laguerre facet between those two neighboring particle polytopes. For construction of the connectivity graph, it is sufficient just to consider joint Laguerre facets between neighboring polytopes of generators in $\mathcal{S}$ since only edges between those generators (vertices) will be part of the connectivity graph.

It is important to mention that in the rest of Section~\ref{sec:model} we consider, e.g., each pair of marked points (generators) or also each pair of neighboring polytopes only once. This means that for all quantities with double indices (e.g., a joint Laguerre facet $F_{nm}$) we only consider the case "$n<m$", where $n,m\in\N$, i.e., we ignore permutations.

Now, based on the tessellation $\mathcal{T}_1$, we place an edge between the generators $(S_n,R_n)$ and $(S_m,R_m)$ whose corresponding polytopes $P_n$ and $P_m$ share a common facet $F_{nm}$ depending on, first, the distance ratio $D_{nm}=\|S_n-S_m\|/(R_n+R_m)$ and, second, the area $A_{nm}=|F_{nm}|$ of $F_{nm}$. Roughly speaking, it is more likely that two particles which will be placed in polytopes induced by $(S_n,R_n)$ and $(S_m,R_m)$ are connected with each other if the distance ratio $D_{nm}$ is small (i.e., close to 1 or smaller) and the area $A_{nm}$ of the common Laguerre facet $F_{nm}$ is large. Note that we do not condition the occurrence of an edge on an angle between the points $S_n$ and $S_m$ like in~\cite{westhoff.2017} as we did not detect any significant anisotropy in the microstructures of scenarios \textsc{P}, \textsc{A1} and \textsc{A2}.

To bring together these thoughts in a (connectivity) graph model, we first construct a random marked graph $G_{all}=(\mathcal{V},\mathcal{E}^\prime)$, where $\mathcal{V}=\mathcal{S}$. The edges $\mathcal{E}^\prime\subset(\mathcal{S}\times\mathcal{S}, \R_+)$ possess non-negative marks $\{P(D_{nm}, A_{nm}), n\neq m\}$, where $P(D_{nm}, A_{nm})$ is the probability that the two vertices $(S_n,R_n)$ and $(S_m,R_m)$ whose corresponding polytopes $P_n$ and $P_m$ in $\mathcal{T}_1$ share a common facet $F_{nm}$ are connected given the distance ratio $D_{nm}$ and the facet area $A_{nm}$. Note that $G_{all}$ contains all edges between pairs of marked points from $\mathcal{S}$ whose corresponding polytopes in $\mathcal{T}_1$ share a common facet. Starting from $G_{all}$, we construct connectivity graph $G=(\mathcal{S},\mathcal{E})$ by adding edges from $\mathcal{E}^\prime$ to the (initially empty) edge set $\mathcal{E}$ according to their probabilities $P(D_{nm}, A_{nm})$. % of each edge $((S_n,R_n),(S_m,R_m),P(D_{nm}, A_{nm}))\in\mathcal{E}^\prime$.
That is, $G$ mainly contains edges from $G_{all}$ having high probability $P(D_{nm}, A_{nm})$. Such connectivity graphs $G$ describe the connectivity relations between particles extracted from our tomographic data quite well, %for example, in all three scenarios the mean coordination numbers are well reproduced (
see Section~\ref{subsec:particlePhaseValid}%)
. In the following, the marks of edges in $\mathcal{E}$ are not necessary anymore and therefore we skip them and just write $((S_n,R_n),(S_m,R_m))\in\mathcal{E}$ instead of $((S_n,R_n),(S_m,R_m),P(D_{nm}, A_{nm}))\in\mathcal{E}$.

\subsubsection{Calibration of connectivity graph model}\label{subsubsec:caliConnGraphModel}

\begin{figure}[!t]
	\centering
	\begin{subfigure}[c]{0.49\textwidth}
		\includegraphics[width=\textwidth]{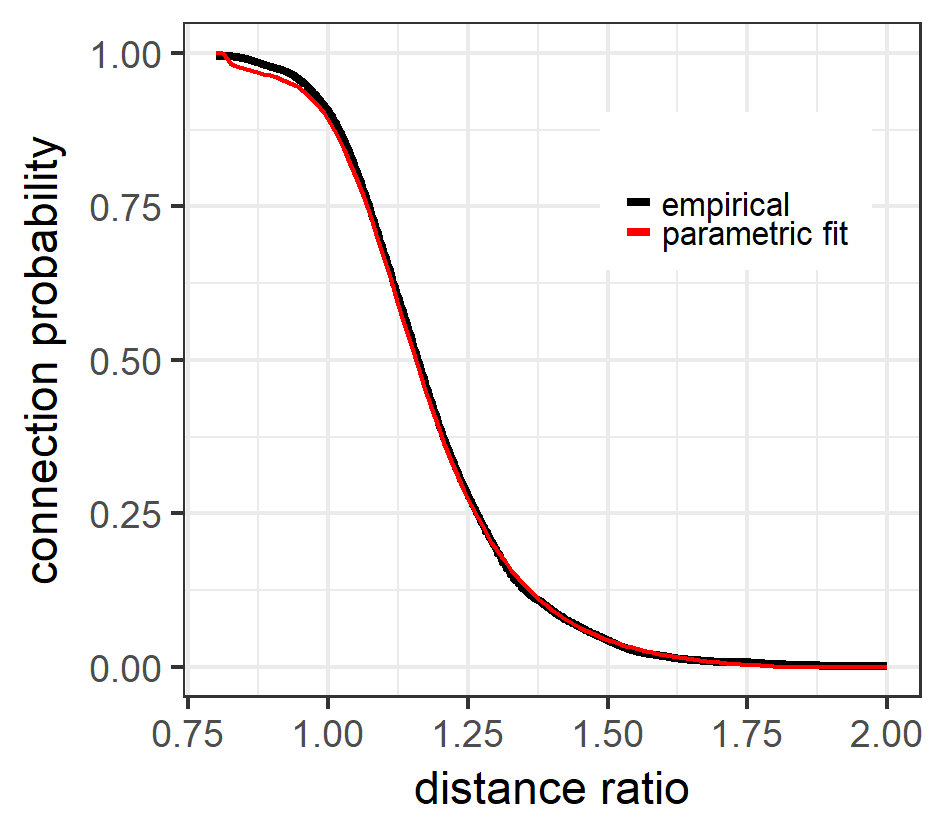}
		\caption{}
		\label{subfig:connectionDist}
	\end{subfigure}
	\begin{subfigure}[c]{0.49\textwidth}
		\includegraphics[width=\textwidth]{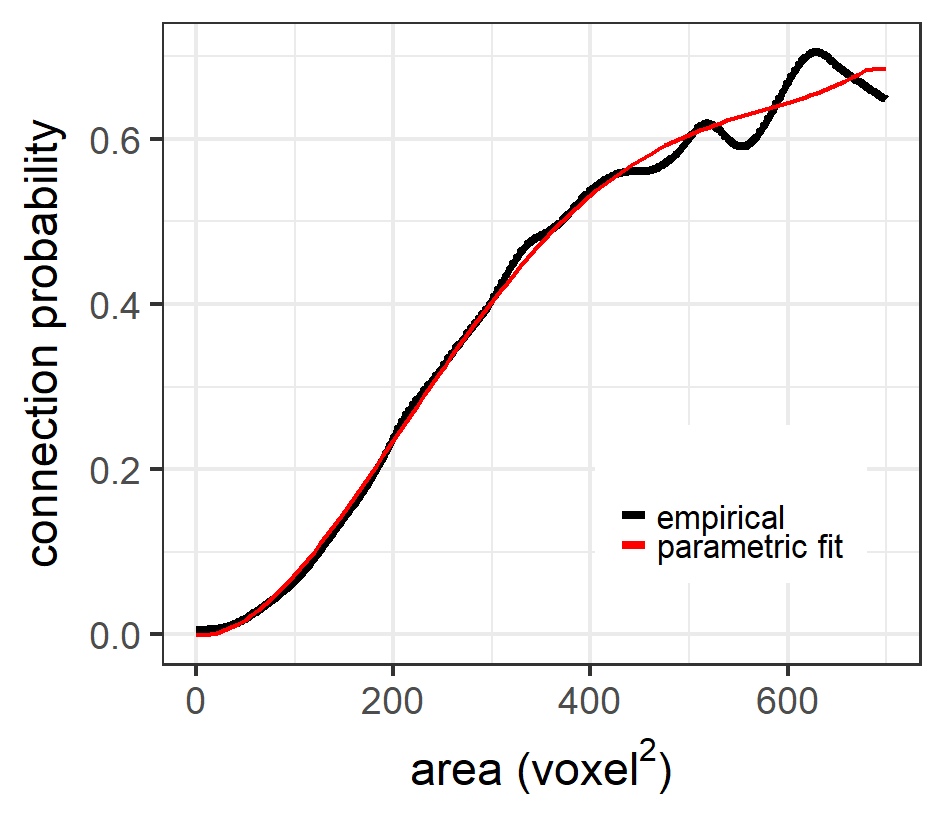}
		\caption{}
		\label{subfig:connectionArea}
	\end{subfigure}
	\caption{(a) Connection probabilities depending on the distance ratio between two particles; (b) Connection probabilities depending on the area of the joint Laguerre facet. Both empirical (connection) probability functions (black curves) are computed for the pristine cathode (scenario \textsc{P}).}
	\label{fig:connection_probability}
\end{figure}

As before for the point pattern model, we have to calibrate the graph model to our tomographic image data. This means that we need the probabilities $P(d,a)$ of two marked points (which represent two particles) being connected in the segmented tomographic microstructures if they have a distance ratio $d>0$ and an area $a>0$ of the common Laguerre facet. Analogous to~\cite{westhoff.2017}, the joint probability $P(d,a)$ is expressed as the product of (marginal) conditional probabilities $P_{dira}(d)$ and $P_{area}(a)$ of two particles being connected given the distance ratio between them is $d$ and the area of the common Laguerre facet is $a$, respectively, multiplied with a correction factor $c>0$. That is, \[P(d,a)=\min\left\{c\cdot P_{dira}(d)\cdot P_{area}(a),\, 1\right\}.\] Both $P_{dira}(d)$ and $P_{area}(a)$ can be estimated as follows. Considering the segmented image data of one of the three scenarios, we know from Section~\ref{subsec:data_prepro} all particles and thus obtain their representation $\{(s^{\scriptscriptstyle{exp}}_n,r^{\scriptscriptstyle{exp}}_n)\}$ as marked point pattern (see the end of Section~\ref{subsubsec:initSphereSystems}), where a particle is located at its center of mass $s^{\scriptscriptstyle{exp}}_n$ and has volume-equivalent radius $r^{\scriptscriptstyle{exp}}_n$. From the image data we also know which pairs of particles are connected. Additionally, we again extract the marked point pattern $\{(\tilde{s}^{\scriptscriptstyle{exp}}_n,\tilde{r}^{\scriptscriptstyle{exp}}_n)\}$ representing the large pores with minimum pore radius greater or equal $t_p$ in exactly the same manner as already described in Section~\ref{subsubsec:initSphereSystems}, where $\tilde{s}^{\scriptscriptstyle{exp}}_n$ denotes the pore location and $\tilde{r}^{\scriptscriptstyle{exp}}_n$ the corresponding minimum pore radius. Then, given these two marked point patterns, we can calculate the Laguerre tessellation $\mathcal{T}^{\scriptscriptstyle{exp}}_1$ induced by the marked point pattern $\{(s^{\scriptscriptstyle{exp}}_n,r^{\scriptscriptstyle{exp}}_n)\}\cup\{(\tilde{s}^{\scriptscriptstyle{exp}}_n,\tilde{r}^{\scriptscriptstyle{exp}}_n)\}$. %$\mathcal{\hat{S}}^{\scriptscriptstyle{exp}}=\mathcal{S}^{\scriptscriptstyle{exp}}\cup\tilde{\mathcal{S}}^{\scriptscriptstyle{exp}}$
Using all this information, we know for each pair of particles represented by $(s^{\scriptscriptstyle{exp}}_n,r^{\scriptscriptstyle{exp}}_n)$ and $(s^{\scriptscriptstyle{exp}}_m,r^{\scriptscriptstyle{exp}}_m)$ if they are connected and if their corresponding convex polytopes in $\mathcal{T}^{\scriptscriptstyle{exp}}_1$ share a common Laguerre facet $f^{\scriptscriptstyle{exp}}_{nm}$. Provided that there exists a common facet $f^{\scriptscriptstyle{exp}}_{nm}$, we can estimate the probability $P_{dira}(d)$ of two particles being connected given the distance ratio between them is $d$. Furthermore, the probability $P_{area}(a)$ of two particles being connected given the area of $f^{\scriptscriptstyle{exp}}_{nm}$ is $a$ can be estimated, see Figure~\ref{fig:connection_probability}.% shows both estimated probabilities (black curves) for the pristine cathode (scenario \textsc{P}) and exhibit the expected behaviors. Note that the behaviors are similar for all scenarios.

For modeling purposes, we fit parametric curves to each estimated probability function $P_{dira}(d)$ and $P_{area}(a)$. That is, we approximate the estimated probability function $P_{dira}(d)$ by \[P_{dira}(d)\approx\
%\begin{cases}
	%1 & \text{if } d<d_{low},\\
	\max\left\{\cfrac{a_{dira}\cdot d+b_{dira}}{d^3+c_{dira}\cdot d^2+d_{dira}\cdot d+e_{dira}},\, 0\right\} %& 
	\text{if } d\geq d_{low},
%\end{cases}
\] where $d_{low}>0$ is a threshold below which $P_{dira}(d)$ is put equal to 1. The estimated probability function $P_{area}(a)$ is approximated by \[P_{area}(a)\approx\min\left\{a_{area}\cdot\hat{a}^4+b_{area}\cdot\hat{a}^3+c_{area}\cdot\hat{a}^2+d_{area}\cdot\hat{a}+e_{area},\, p_{cap}\right\},\] where $\hat{a}=(a-m_{area})/s_{area}$ is some normalization of $a$. All parameters of the approximations of $P_{dira}(d)$ and $P_{area}(a)$ have been determined using the curve fitting toolbox in MATLAB~\cite{matlab.2016b} and their values for the scenarios \textsc{P}, \textsc{A1} and \textsc{A2} can be found in Table~\ref{tab:graphParam}. For scenario \textsc{P} the fitted parametric curves are shown in Figure~\ref{fig:connection_probability} (red curves).

\begin{table}[ht!]
	\centering
	\begin{tabular}{r|c|c|c}
		\hline
		& \multicolumn{3}{c}{scenario}\\
		& \textsc{P} & \textsc{A1} & \textsc{A2}\\
		\hline
		$a_{dira}$ & -0.04 & -0.03 & -0.03\\
		$b_{dira}$ & 0.06 & 0.05 & 0.06\\
		$c_{dira}$ & -2.65 & -2.68 & -2.63\\
		$d_{dira}$ & 2.31 & 2.36 & 2.28\\
		$e_{dira}$ & -0.63 & -0.66 & -0.62\\
		$d_{low}$ & 0.814 & 0.815 & 0.803\\
		\hdashline
		$a_{area}$ & 0.02 & 0.03 & 0.04\\
		$b_{area}$ & -0.02 & -0.008 & -0.003\\
		$c_{area}$ & -0.11 & -0.13 & -0.16\\
		$d_{area}$ & 0.26 & 0.28 & 0.26\\
		$e_{area}$ & 0.47 & 0.52 & 0.55\\
		$m_{area}$ & 350 & 350 & 317.5\\
		$s_{area}$ & 202.7 & 202.7 & 183.8\\
		$p_{cap}$ & 0.685 & 0.8 & 0.815\\
		\hdashline
		$c$ & 1.7875 & 1.5570 & 1.4824\\
		\hline
	\end{tabular}
	\caption{Fitted curve parameters for the function approximations of $P_{dira}(d)$ and $P_{area}(a)$, listed for each tomographic data set of all three scenarios \textsc{P}, \textsc{A1} and \textsc{A2}. Each optimized correction factor $c$ is also listed.}
	\label{tab:graphParam}
\end{table}

To determine the correction factor $c$ in the definition of $P(d,a)$ we apply exactly the same procedure, namely the minimum contrast method, as performed in~\cite{westhoff.2017}. Thus, the cost function \[h(c)=|\kappa^{\scriptscriptstyle{exp}}-\kappa(c)|\] is minimized, where $\kappa^{\scriptscriptstyle{exp}}$ is the mean coordination number observed in the connectivity graph of a (segmented) tomographic data set and $\kappa(c)$ is the mean coordination number observed in connectivity graphs which are realized by the model given the correction factor is equal to $c$. Note that the mean coordination number is the mean number of edges emanating from a vertex and for the three scenarios \textsc{P}, \textsc{A1} and \textsc{A2} we found the rounded values 2.69, 3.15 and 3.30 for $\kappa^{\scriptscriptstyle{exp}}$. The value of $\kappa(c)$ is determined by generating 100 realizations of the model (i.e., point pattern model plus graph model with given value of $c$) and then we average over all mean coordination numbers observed in these realizations. By the bisection method~\cite{burden.2010}, the minimization of the cost function $h$ is executed starting from a suitable initial interval for possible values of $c$. The optimized correction factor for each scenario can be found in Table~\ref{tab:graphParam}.

\subsection{Modeling of suitable particle polytopes and contact conditions}\label{subsec:tessModel}

Up to this point, the stochastic 3D microstructure model involves the generation of two random marked point patterns which approximately configure the large pores and the particles of the cathode microstructure. Furthermore, the model uses a Laguerre tessellation to determine the connectivity between particles via a random graph. But the Laguerre tessellation or, more precisely, its corresponding collection of convex polytopes has two further functionalities, namely, it influences sizes and shapes of particles and it also sets contact conditions on polytope boundaries which have to be fulfilled.

\subsubsection{Enhanced insertion of pore polytopes}\label{subsubsec:porePoly}% due to low volume fraction}\label{subsubsec:emptyPoly}

To achieve desired sizes and shapes of particles, we have to struggle with the same kind of modeling issue as in~\cite{westhoff.2017}. The difficulty is that, due to the low volume fraction of the particle phase, the particle polytopes $\{P_n\}\subset\mathcal{T}_1$ from Section~\ref{subsubsec:particleConnections} are currently too large and badly shaped in order to achieve reasonably particles under the connectivity constraints prescribed by the graph $G$. Preferably, we need particle polytopes which have the same order of size as the particles being placed into and possess a nearly spherical shape similar to, e.g., regular icosahedra or dodecahedra. For this purpose, we enhance the idea considered in~\cite{westhoff.2017} of adding (further) pore polytopes which remain empty. The intention of adding further pore polytopes is to gain suitable (i.e., smaller and more spherical) particle polytopes. Before we explain the rather technical procedure how further pore polytopes are added, we introduce some helpful notation.

Let $\mathcal{F}=\{F_{nm}, ((S_n,R_n),(S_m,R_m))\in\mathcal{E}\}$ denote the joint Laguerre facets $F_{nm}$ between pairs of neighboring particle polytopes $P_n$ and $P_m$, where the connectivity graph $G$ from Section~\ref{subsec:graphModel} indicates a connection via this facet. Furthermore, let $\mathcal{C}=\{C_{nm}\}$ denote the centroids of these facets. Similar to $\mathcal{F}$, the set $\mathcal{F}^\prime=\{F_{nm}, ((S_n,R_n),(S_m,R_m))\notin\mathcal{E}\}$ contains joint facets between pairs of particle polytopes, where $G$ does not indicate a connection. The set of centroids of facets in $\mathcal{F}^\prime$ is denoted by $\mathcal{C}^\prime=\{C_{nm}^\prime\}$. A third set of facets $\mathcal{F}^{\prime\prime}=\{\tilde{F}_{nm}\}$ includes joint facets $\tilde{F}_{nm}$ between a particle polytope $P_n$ and a pore polytope $\tilde{P}_m$. The corresponding set of facet centroids is denoted by $\mathcal{C}^{\prime\prime}=\{C_{nm}^{\prime\prime}\}$. Finally, we introduce the set of polytope vertices $\{V_l\}$ of all particle polytopes $\{P_n\}$. Note that most of the $V_l$ belong to more than one particle polytope but are considered within the set $\{V_l\}$ only once.

\begin{figure}[!ht]
	\centering
	\begin{subfigure}[c]{0.49\textwidth}
		\includegraphics[width=\textwidth]{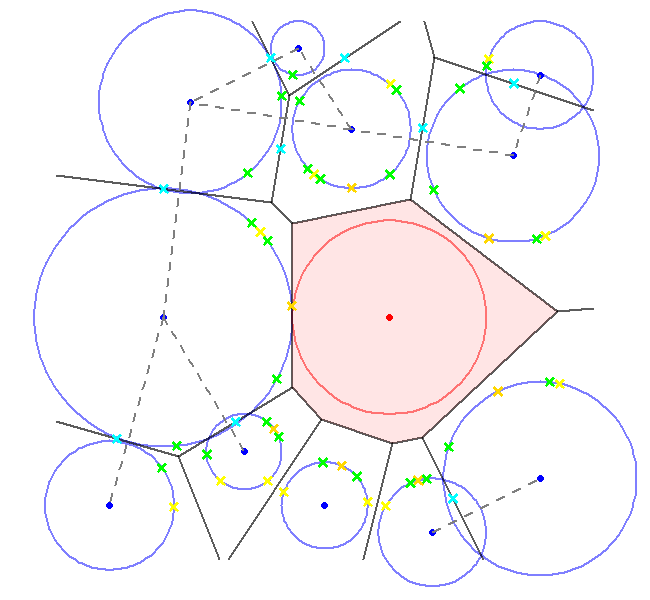}
		\caption{}
		\label{subfig:restrictions}
	\end{subfigure}
	\begin{subfigure}[c]{0.49\textwidth}
		\includegraphics[width=\textwidth]{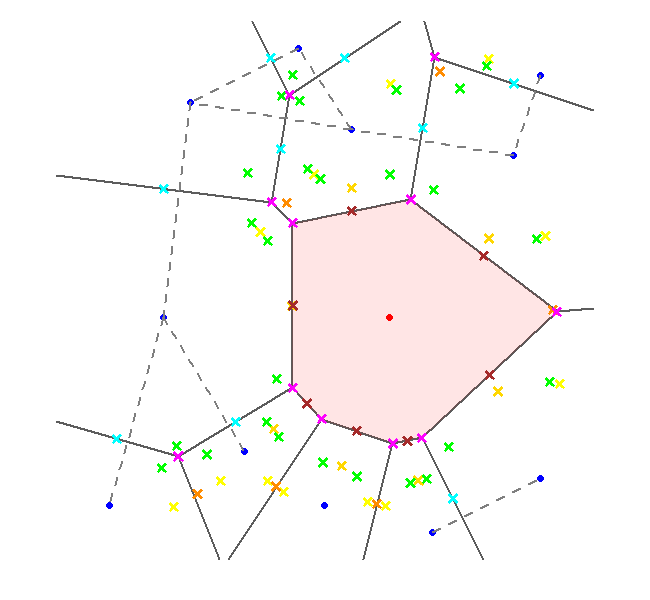}
		\caption{}
		\label{subfig:restrAndCand}
	\end{subfigure}
	\caption{Enhanced insertion of pore polytopes: (a) Initial tessellation $\mathcal{T}_1$ with red shaded large pore polytope. The restriction points $\mathcal{R}$ (cyan, yellow, golden and green crosses) may not be covered by further added pore polytopes since they ensure the desired connectivity of particles and lead to suitable particle polytopes in the end; (b) Candidates for generators of further pore polytopes $\hat{\mathcal{S}}_{cand}$ (orange, brown and magenta crosses) are determined. Each candidate point possesses three descending marks and therefore it also has three chances to generate an additional further pore polytope.}
	\label{fig:porePolySketch1}
\end{figure}

Based on this notation, we can explain the procedure how further pore polytopes are added. Since each convex polytope of the tessellation $\mathcal{T}_1$ is induced by a marked point, i.e., by its generator, in $\mathcal{S}\cup\tilde{\mathcal{S}}$, we try to find candidates for additional generators of pore polytopes which shrink the particle polytopes $\{P_n\}$ to a suitable size and shape. However, the shrinkage of particle polytopes is restricted in such a way that it does not violate two requirements. First, we ensure the persistent existence of each facet $F_{nm}\in\mathcal{F}$ through the requirement that no corresponding centroid $C_{nm}\in\mathcal{C}$ (see Figure~\ref{subfig:restrictions}, cyan crosses) is covered by a pore polytope of a possibly additional generator. Note that this first requirement guarantees reasonable connectivity according to the graph $G$.
%\textcolor{red}{TODO: Position of "hardShell-points", i.e., on the surface or not!?}
Second, we introduce a set of points $\mathcal{R}$ on the surfaces of the (generator) spheres $B(S_n,R_n)$, see Figure~\ref{subfig:restrictions} blue circles, inducing the particle polytopes $P_n$. We also require that no point of $\mathcal{R}$ is covered by a pore polytope of a candidate generator. Note that this second requirement prevents too much shrunken particle polytopes. So $\mathcal{R}$ consists of the following points:
\begin{enumerate}[nolistsep, leftmargin=20pt]
\item[(a)] Each facet centroid $C_{nm}^\prime\in\mathcal{C}^\prime$ is associated with two neighboring particle polytopes induced by their generator spheres $B(S_n,R_n)$ and $B(S_m,R_m)$. Let $\vec{V}_{n}^\prime$ be the vector from $S_n$ to $C_{nm}^\prime$ scaled to length $R_n$ and $\vec{V}_{m}^\prime$ the vector from $S_m$ to $C_{nm}^\prime$ scaled to length $R_m$. Then, $\mathcal{R}$ contains the two points $S_n+\vec{V}_{n}^\prime$ and $S_m+\vec{V}_{m}^\prime$ which belong to the surfaces of $B(S_n,R_n)$ and $B(S_m,R_m)$, respectively (see Figure~\ref{subfig:restrictions}, yellow crosses);
\item[(b)] Each facet centroid $C_{nm}^{\prime\prime}\in\mathcal{C}^{\prime\prime}$ is associated with a particle polytope induced by its generator sphere $B(S_n,R_n)$. Let $\vec{V}_{n}^{\prime\prime}$ be the vector from $S_n$ to $C_{nm}^{\prime\prime}$ scaled to length $R_n$. Then, $\mathcal{R}$ also contains the point $S_n+\vec{V}_{n}^{\prime\prime}$ which belongs to the surface of $B(S_n,R_n)$ (see Figure~\ref{subfig:restrictions}, golden crosses);
\item[(c)] Each particle polytope vertex $V_l$ is associated with up to four particle polytopes induced by the generator spheres $B(S_{l_1},R_{l_1}), B(S_{l_2},R_{l_2}),\dots~$. Let $\vec{V}_{l_1}$ be the vector from $S_{l_1}$ to $V_l$ scaled to length $R_{l_1}$, $\vec{V}_{l_2}$ the vector from $S_{l_2}$ to $V_l$ scaled to length $R_{l_2}$ and so on. Then, $\mathcal{R}$ additionally contains the points $S_{l_1}+\vec{V}_{l_1}, S_{l_2}+\vec{V}_{l_2},\dots$ which belong to the surfaces of $B(S_{l_1},R_{l_1}), B(S_{l_2},R_{l_2}),\dots$, respectively (see Figure~\ref{subfig:restrictions}, green crosses).
\end{enumerate}
Finally, we add to $\mathcal{R}$ all centroids $C_{nm}$ which are relevant to the first requirement and for practical reasons we consistently denote the points of $\mathcal{R}$ by $X_k$., i.e., $\mathcal{R}=\{X_k, k\in\N\}$. Thus, $\mathcal{R}$ contains all points which may not be covered and hence imply a restriction to adding further pore polytopes.

Next, we define the set $\hat{\mathcal{S}}_{cand}$ of candidates for additional generators of pore polytopes. The first candidates we add to $\hat{\mathcal{S}}_{cand}$ are modifications of the centroids in $\mathcal{C}^\prime$ of the joint facets between two neighboring particle polytopes induced by their generator spheres $B(S_n,R_n)$ and $B(S_m,R_m)$. 
Recall that on the surface of each of these two spheres we have defined a point which is contained in $\mathcal{R}$ (see item (a) above). Then, we shift the point $C_{nm}^\prime\in\mathcal{C}^\prime$ in such a way that it is placed in the middle between its two corresponding points from $\mathcal{R}$. Doing the same for all $C_{nm}^\prime$, we add the resulting (shifted) points to $\hat{\mathcal{S}}_{cand}$ (see Figure~\ref{subfig:restrAndCand}, orange crosses). Finally, we complement the set $\hat{\mathcal{S}}_{cand}$ by adding all centroids $C_{nm}^{\prime\prime}$ (see Figure~\ref{subfig:restrAndCand}, brown crosses) and all particle polytope vertices $V_l$ (see Figure~\ref{subfig:restrAndCand}, magenta crosses) as candidates for additional generators of pore polytopes.

\begin{figure}[!pht]
	\centering
	\begin{subfigure}[c]{0.49\textwidth}
		\includegraphics[width=\textwidth]{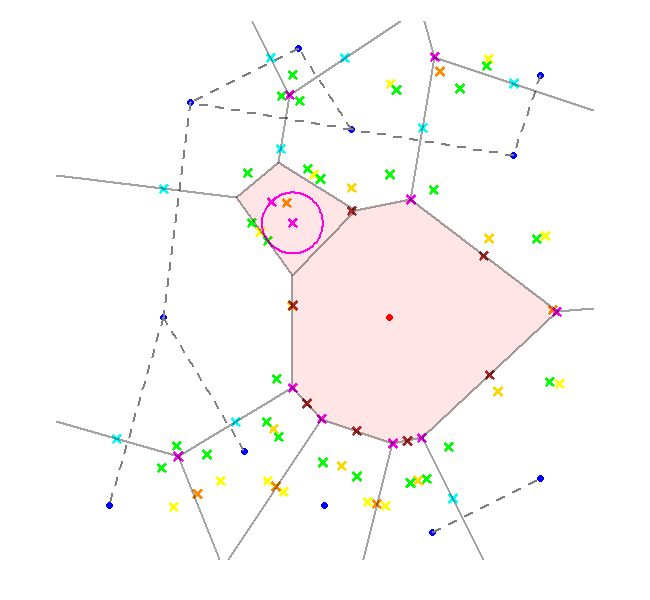}
		\caption{}
		\label{subfig:acceptedPorePoly}
	\end{subfigure}
	\begin{subfigure}[c]{0.49\textwidth}
		\includegraphics[width=\textwidth]{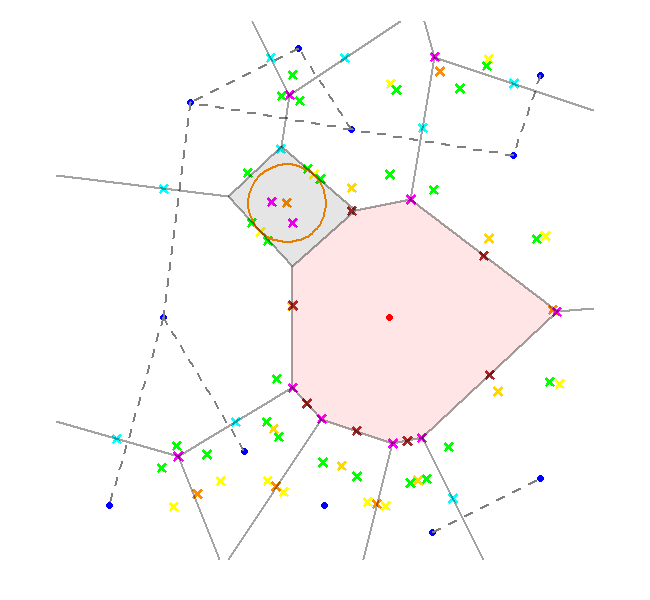}
		\caption{}
		\label{subfig:firstRejected}
	\end{subfigure}
	\begin{subfigure}[c]{0.49\textwidth}
		\includegraphics[width=\textwidth]{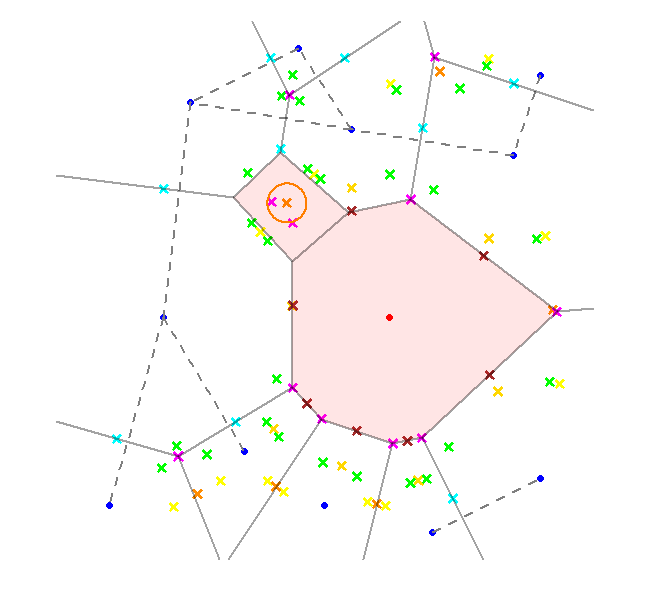}
		\caption{}
		\label{subfig:thenAccepted}
	\end{subfigure}
	\begin{subfigure}[c]{0.49\textwidth}
		\includegraphics[width=\textwidth]{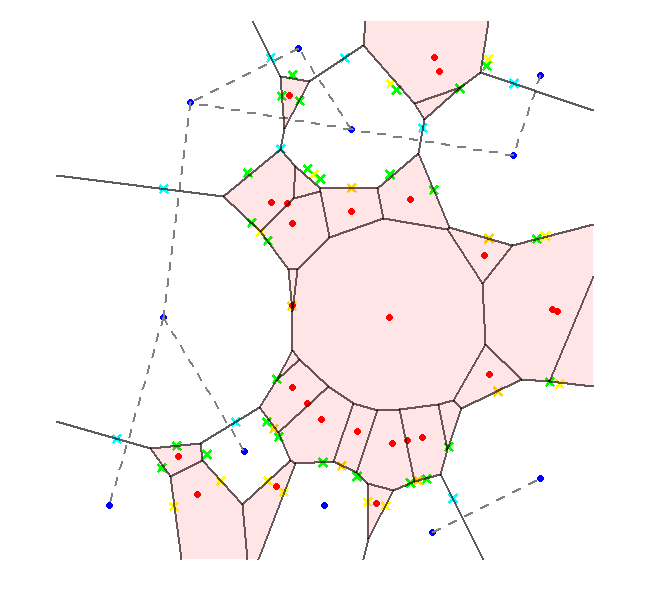}
		\caption{}
		\label{subfig:restrAndTessFinal}
	\end{subfigure}
	\caption{Enhanced insertion of pore polytopes: (a) The small red shaded polytope induced by the current candidate generator $(\hat{S}_n,\hat{R}^{(1)}_n)\in\hat{\mathcal{S}}_{cand}$ (magenta cross with circle) does not cover any of the restriction points $X_k\in\mathcal{R}$ (cyan, yellow, golden or green crosses), therefore this generator is accepted and added to $\hat{\mathcal{S}}$; (b) The next candidate $(\hat{S}_m,\hat{R}^{(1)}_m)$ (orange cross with circle) induces the gray shaded polytope which covers a restriction point $X_k$ (cyan cross in the upper corner), therefore $(\hat{S}_m,\hat{R}^{(1)}_m)$ is rejected; (c) In this case, we test the previous candidate with the smaller mark $\hat{R}^{(2)}_m$ (orange cross with circle) and the induced polytope covers no point of $\mathcal{R}$, therefore $(\hat{S}_m,\hat{R}^{(2)}_m)$ is accepted and added to $\hat{\mathcal{S}}$; (d) Final tessellation (pore polytopes are red shaded with generators $\tilde{\mathcal{S}}\cup\hat{\mathcal{S}}$ as red dots) induced by the marked point pattern $\mathcal{S}\cup\tilde{\mathcal{S}}\cup\hat{\mathcal{S}}$.}
	\label{fig:porePolySketch2}
\end{figure}

So far, the candidates in $\hat{\mathcal{S}}_{cand}$ possess no marks to serve as generators for polytopes in a Laguerre tessellation. Therefore, a mark (radius) $\hat{R}^{(i)}_n>0$ is assigned to each point in $\hat{\mathcal{S}}_{cand}$. From now on, we consistently denote the candidates for generators in $\hat{\mathcal{S}}_{cand}$ by $(\hat{S}_n,\hat{R}^{(i)}_n)$ for $i\in\{1,2,3\}$. That is, every point $\hat{S}_n$ could be accepted under three different constellations as an additional generator for a pore polytope. The primary constellation is $(\hat{S}_n,\hat{R}^{(1)}_n)$, where $\hat{R}^{(1)}_n$ is equal to 0.99 times the distance from $\hat{S}_n$ to the nearest point $X_k\in\mathcal{R}$. The second constellation is $(\hat{S}_n,\hat{R}^{(2)}_n)$, where $\hat{R}^{(2)}_n=\hat{R}^{(1)}_n/2$ and the last constellation to test is $(\hat{S}_n,\hat{R}^{(3)}_n)$, where $\hat{R}^{(3)}_n=0$.

Given the candidates for generators in $\hat{\mathcal{S}}_{cand}$ and the points in $\mathcal{R}$, we thin out the set $\hat{\mathcal{S}}_{cand}$ such that no point $X_k\in\mathcal{R}$ is covered by a pore polytope induced by a candidate generator $(\hat{S}_n,\hat{R}^{(i)}_n)\in\hat{\mathcal{S}}_{cand}$. This can be checked as follows. For each candidate generator $(\hat{S}_n,\hat{R}^{(i)}_n)$, we start with the primary constellation $(\hat{S}_n,\hat{R}^{(1)}_n)$ and consider the Laguerre tessellation $\mathcal{T}_{(\hat{S}_n,\hat{R}^{(1)}_n)}$ induced by the existing marked point pattern $\mathcal{S}\cup\tilde{\mathcal{S}}$, additionally including $(\hat{S}_n,\hat{R}^{(1)}_n)$. If the pore polytope $\hat{P}_n\in\mathcal{T}_{(\hat{S}_n,\hat{R}^{(1)}_n)}$ of generator $(\hat{S}_n,\hat{R}^{(1)}_n)$ does not cover any point $X_k\in\mathcal{R}$, then we add this generator to the final set $\hat{\mathcal{S}}$ of additional generators of pore polytopes (see Figure~\ref{subfig:acceptedPorePoly}). Otherwise, e.g., shown in Figure~\ref{subfig:firstRejected}, we do the same check for the second constellation $(\hat{S}_n,\hat{R}^{(2)}_n)$, see Figure~\ref{subfig:thenAccepted}, and, if necessary, also for the third one with mark equal to zero. Should all three constellations be rejected, then this candidate generator $(\hat{S}_n,\hat{R}^{(i)}_n)$ for $i\in\{1,2,3\}$ is not accepted and not added to $\hat{\mathcal{S}}$.

The final tessellation $\mathcal{T}$, which is induced by the marked point pattern $\mathcal{S}\cup\tilde{\mathcal{S}}\cup\hat{\mathcal{S}}$, has suitable particle polytopes $\{P_n\}\subset\mathcal{T}$, i.e., the particle polytopes $P_n$ have desired sizes and shapes in order to achieve reasonable particles under the connectivity constraints given by the graph $G$. This final result is sketched in Figure~\ref{subfig:restrAndTessFinal}.

\subsubsection{Setting of suitable contact conditions}\label{subsubsec:connCond}% due to more spherical particles}\label{subsubsec:connCond}

In this section we define the conditions which particles placed into polytopes have to fulfill in order to ensure connectivity imposed by the graph $G$ from Section~\ref{subsec:graphModel}. Recall that the particles shall be connected via joint Laguerre facets of their neighboring particle polytopes, where $G=(\mathcal{S},\mathcal{E})$ indicates connections through edges in $\mathcal{E}$. Therefore, we will force the particles to cover some specific points on such facets of their polytopes (see Section~\ref{subsubsec:particleCreation}).

Recall that the Laguerre facets of the tessellation $\mathcal{T}$ where particles are supposed to be connected are those of $\mathcal{F}$. The corresponding set of facet centroids is given by $\mathcal{C}$, see Section~\ref{subsubsec:porePoly}. Thus, we denote the set of points which a particle being placed into the polytope $P_n$ has to cover by $\mathcal{C}_n=\{C_{nm}\in\mathcal{C}, n<m, m\in\N\}\cup\{C_{ln}\in\mathcal{C}, l<n, l\in\N\}$ for each $n\in\N$. Furthermore, for convenience, we denote the $N<\infty$ points in $\mathcal{C}_n$ by $C_i$, i.e., $\mathcal{C}_n=\{C_i, i=1,\dots,N\}$. Consequently, each pair of connected particles will be at least connected in the centroid of the joint facet between their particle polytopes, see light blue dots in Figure~\ref{subfig:tessFinalAndParticles}. This, so to say, one-point-per-facet contact condition has two advantages for the creation of particles. On the one hand, we tendentially achieve small contact regions between particles similar to those observed in tomographic data. On the other hand, the small total number of contact conditions per particle helps to create the more spherical-shaped cathode particles.

\subsection{Creation of particles under contact conditions}\label{subsec:particleModel}

In the previous modeling steps we have prepared the basis which now puts us in a position to create the particles themselves. The particles, like in~\cite{feinauer.2015b} and~\cite{westhoff.2017}, are modeled as realizations of (conditional) Gaussian random fields on the sphere (GRF for short) which are expressed in terms of spherical harmonics series expansions. %Since this last modeling step sometimes does not differ in its ideas from those introduced in~\cite{feinauer.2015b} and~\cite{westhoff.2017}, we partially just recapitulate the most important aspects.

\subsubsection{Representation of particles as spherical harmonics series expansion}\label{subsubsec:spherHarm}

As just mentioned, each particle is created as a realization of a so-called (isotropic) Gaussian random field on the sphere $\psi\colon[0,\pi]\times[0,2\pi)\to\R$ under the given (contact) conditions from Section~\ref{subsubsec:connCond}. %, therefore one can speak of a conditional Gaussian random field.
Then, the extension of a random particle in each direction $(\theta,\phi)$ with respect to some center point is given by the value $\psi(\theta,\phi)$ of the GRF in this direction. Note that the GRF $\psi$ can be uniquely described by a mean radius $\mu\in\R$ and a so-called angular power function $A\colon[0,\infty)\to[0,\infty)$ and has the advantage that it can be expressed in terms of a spherical harmonics series expansion~\cite{lang.2015}. Part of such a series expansion are the so-called spherical harmonics $Y_{l,m}\colon[0,\pi]\times[0,2\pi)\to\C$ for $l\in\N_0$ and $m\in\{0,\dots,l\}$ which are given by \[Y_{l,m}(\theta,\phi)=\sqrt{\frac{(2l+1)}{4\pi}\frac{(l-m)!}{(l+m)!}}P_{l,m}(\cos(\theta))e^{im\phi},\] where the associated Legendre functions $P_{l,m}\colon[-1,1]\to\R$ are defined as \[P_{l,m}(x)=\frac{(-1)^m(1-x^2)^{m/2}}{2^l\, l!}\frac{\mathrm{d}^{l+m}}{\mathrm{d}x^{l+m}}(x^2-1)^l.\] Then, the corresponding series expansion of a GRF with mean radius $\mu$ and angular power function $A$ has the form
\begin{align*}
\psi(\theta,\phi)=\sum_{l=0}^\infty\Bigl[a_{l,0}Y_{l,0}(\theta,\phi)+2\sum_{m=1}^l\bigl[&\mathrm{Re}(a_{l,m})\mathrm{Re}(Y_{l,m}(\theta,\phi))\\
 &-\mathrm{Im}(a_{l,m})\mathrm{Im}(Y_{l,m}(\theta,\phi))\bigr]\Bigr].
\end{align*}
The random coefficients $a_{l,m}, \mathrm{Re}(a_{l,m})$ and $\mathrm{Im}(a_{l,m})$ possess normal distributions, namely, $a_{0,0}\sim\mathcal{N}(\mu/Y_{0,0}(\theta,\phi),A_0)$ with $Y_{0,0}(\theta,\phi)=\frac{1}{2\sqrt{\pi}}$, $a_{l,0}\sim\mathcal{N}(0,A_l)$ for $l>0$, $\mathrm{Re}(a_{l,m})\sim\mathcal{N}(0,A_l/2)$ and $\mathrm{Im}(a_{l,m})\sim\mathcal{N}(0,A_l/2)$ for $l\in\N, m\in\{1,\dots,l\}$, where $\{A_l, l=0,1,\dots\}$ with $A_l=A(l)$ is called the angular power spectrum. For numerical computations the series expansion of $\psi$ has to be truncated at a certain (finite) degree of accuracy $L<\infty$, i.e., the outer sum extends from 0 to $L$ (as opposed to infinity). %GRF is approximated by 
%\begin{align*}
%\psi(\theta,\phi)\approx\sum_{l=0}^L\Bigl[a_{l,0}Y_{l,0}(\theta,\phi)+2\sum_{m=1}^l\bigl[&\mathrm{Re}(a_{l,m})\mathrm{Re}(Y_{l,m}(\theta,\phi))\\
 %&-\mathrm{Im}(a_{l,m})\mathrm{Im}(Y_{l,m}(\theta,\phi))\bigr]\Bigr].
%\end{align*}
To generate such an approximation of the GRF, all we have to do is to sample a realization of the normal distributed random vector $(a_{0,0},\allowbreak a_{1,0},\allowbreak \mathrm{Re}(a_{1,1}),\allowbreak \mathrm{Im}(a_{1,1}),\allowbreak\dots,\allowbreak \mathrm{Re}(a_{L,L}),\allowbreak \mathrm{Im}(a_{L,L}))^{\mathsf T}$. Actually, we will sample from a conditional multivariate normal distribution, see Section~\ref{subsubsec:particleCreation}, since we want each particle (being a GRF) to fulfill given contact conditions. But before we can create individual particles via a GRF, we need to determine the parameters of the (conditional) multivariate normal distribution, i.e., the values of the (truncated) angular power spectrum $\{A_l, l=0,\dots,L\}$ and the mean radius of each particle. Note that the truncation parameter $L$ as degree of accuracy will be chosen dynamically for each particle (see Section~\ref{subsubsec:dynamicL} below) in distinction to the fixed value used in~\cite{feinauer.2015b} and~\cite{westhoff.2017}.

\subsubsection{Adjustment of angular power function and mean radius}\label{subsubsec:aps}

In this section, we determine the parameters of the multivariate normal distributions mentioned in Section~\ref{subsubsec:spherHarm}. For this purpose, we begin to estimate the values of the angular power spectrum from the segmented tomographic image data. To do so, we describe each segmented particle by a spherical harmonics series expansion truncated at $L=20$, see~\cite{feinauer.2015a} for details, and thus we know the series coefficients describing each segmented particle. Given these coefficients, we can estimate their variances and in this way we also obtain estimates of the values $A_l$ for $l=1,\dots,L$. To these data points $(l,A_l)$ we fit a parametric function, see Figure~\ref{fig:apsP}, which provides an appropriate approximation of the angular power function, i.e., \[A(l)\approx\frac{a_A\cdot l+b_A}{l^2+c_A\cdot l+d_A}.\] By the curve fitting toolbox in MATLAB we get the values of $a_A$, $b_A$, $c_A$ and $d_A$ for the scenarios \textsc{P}, \textsc{A1} and \textsc{A2} as listed in Table~\ref{tab:apf}.
\begin{table}[ht]
  \begin{minipage}[b]{0.475\textwidth}
    \centering
    \includegraphics[scale=0.52]{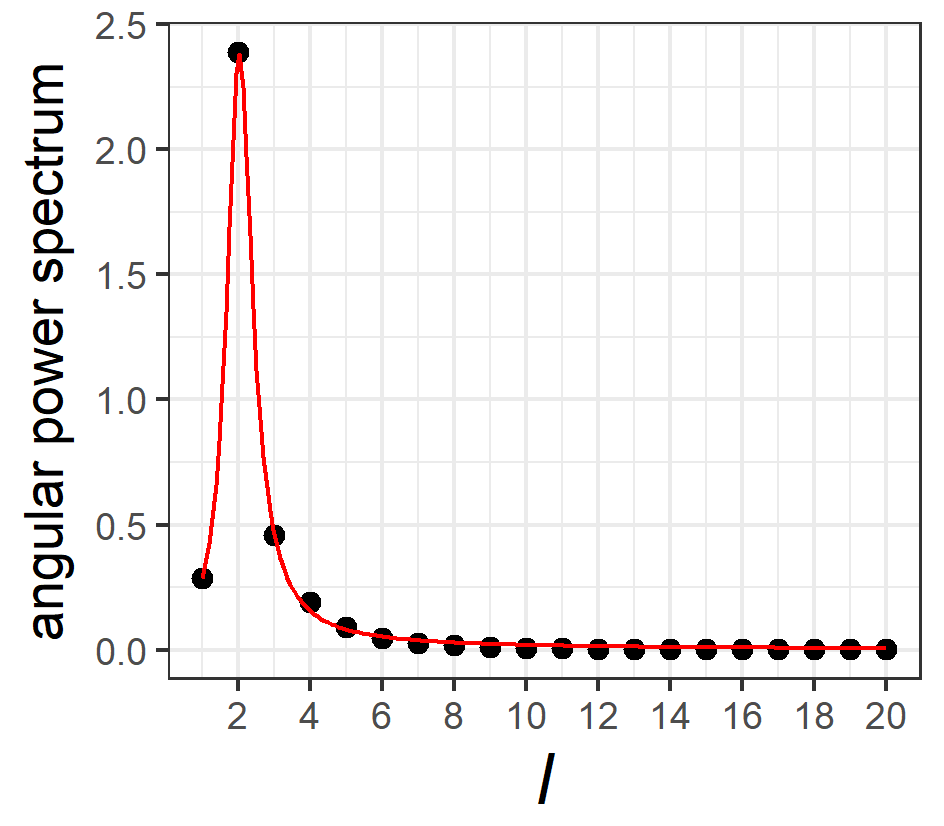}
    \captionof{figure}{Fitted angular power function (red) to the estimated angular power spectrum (black dots) for scenario \textsc{P}.}
    \label{fig:apsP}
  \end{minipage}
	\hfill
	\begin{minipage}[b]{0.475\textwidth}
    \centering
    \begin{tabular}{r|c|c|c}
      \hline
			& \multicolumn{3}{c}{scenario}\\
      & \textsc{P} & \textsc{A1} & \textsc{A2}\\
      \hline
      $a_A$ & 0.1027 & 0.1217 & 0.1049\\
      $b_A$ & 0.2411 & 0.1892 & 0.1066\\
      $c_A$ & -4.009 & -4.019 & -4.064\\
      $d_A$ & 4.206 & 4.215 & 4.284\\
      \hline
			\multicolumn{4}{c}{}\\
    \end{tabular}
    \caption{Fitted values of parameters $a_A$, $b_A$, $c_A$ and $d_A$.\newline}
    \label{tab:apf}
  \end{minipage}
\end{table}

Recall that the mean radius $\mu$ of each random particle is mainly controlled by the coefficient $a_{0,0}$ (it holds $a_{0,0}\sim\mathcal{N}(\mu/Y_{0,0}(\theta,\phi),A_0)$, whereas all other coefficients have expectation zero). Thus, to control the expected volume $\E V=\frac{1}{3}\int_0^{2\pi}{\int_0^\pi{\E[\psi(\theta,\phi)^3]\sin(\theta)\mathrm{d}\theta}\mathrm{d}\phi}$ of each random particle, we choose the coefficient $a_{0,0}$ deterministically depending on each particle polytope $P_n\in\mathcal{T}$ and the volume fraction $\nu^{\scriptscriptstyle{exp}}/\xi$ which a particle should cover being placed into $P_n$. Here, $\nu^{\scriptscriptstyle{exp}}$ is the volume fraction of the particle phase observed in a tomographic data set and $\xi$ is the volume fraction of all particle polytopes $\{P_n\}\subset\mathcal{T}$ obtained in a model realization. The values of $\nu^{\scriptscriptstyle{exp}}$ for the three scenarios \textsc{P}, \textsc{A1} and \textsc{A2} can be found in Table~\ref{tab:volSSA}. Under the assumption of independence of all random series coefficients $a_{l,m}, \mathrm{Re}(a_{l,m})$ and $\mathrm{Im}(a_{l,m})$, it is possible to show that the expected volume $\E V$ of a particle has the form \[\E V=|B(o,1/2\sqrt{\pi})|\, a_{0,0}^3+\Sigma_{L}\, a_{0,0},\] where $|B(o,\frac{1}{2\sqrt{\pi}})|\approx 0.09403$ is the volume of a sphere with radius $\frac{1}{2\sqrt{\pi}}$ around the origin $o$. The parameter $\Sigma_{L}$ can be written as
\begin{align*}
\Sigma_{L}=\frac{1}{2\sqrt{\pi}}\sum_{l=1}^LA_l\int_0^{2\pi}\int_0^\pi\Bigl[&Y_{l,0}(\theta,\phi)^2+2\sum_{m=1}^l\bigl[\mathrm{Re}(Y_{l,m}(\theta,\phi))^2\\
 &+\mathrm{Im}(Y_{l,m}(\theta,\phi))^2\bigr]\Bigr]\sin(\theta)\mathrm{d}\theta\mathrm{d}\phi.
\end{align*}
It depends on $L$ and the estimates of the angular power spectrum $A_l$ for $l=1,\dots,L$. This means that $\Sigma_{L}$ can vary from particle to particle since $L$ is chosen dynamically. Then, for each particle polytope $P_n$ with volume $|P_n|$ we get the coefficient $a_{0,0}$ by solving the equation (which has a unique real-valued solution) \[\rho|P_n|\nu^{\scriptscriptstyle{exp}}/\xi=0.09403a_{0,0}^3+\Sigma_{L}\, a_{0,0}.\] As in~\cite{westhoff.2017}, $\rho$ is a correction factor which compensates for the errors that might occur due to the involvement of contact conditions $\mathcal{C}_n$ from Section~\ref{subsubsec:connCond}. This means that the assumption of independent series coefficients made above is no longer valid under given contact conditions. Therefore, we compensate for this missing assumption by adding the factor $\rho$. Note that the correction factor $\rho$ is determined as a the root of $h(\rho)=|\nu^{\scriptscriptstyle{exp}}-\nu(\rho)|$ using the bisection method, where $\nu(\rho)$ is the mean volume fraction of the particle phase over 10 model realizations with the given $\rho$. In this way, for the scenarios \textsc{P}, \textsc{A1} and \textsc{A2}, we get the values $\rho_{\textsc{P}}=0.9891$, $\rho_{\textsc{A1}}=0.9834$ and $\rho_{\textsc{A2}}=0.9644$.

Finally, if we choose the coefficient $a_{0,0}$ for each particle in such a way, it ensures us to generate particle systems with the desired volume fraction $\nu^{\scriptscriptstyle{exp}}$ on average.

\subsubsection{Relate degree of accuracy L to coordination number}\label{subsubsec:dynamicL}

To account for the typically spherical\hyp{}shaped cathode particles, we choose the degree of accuracy $L<\infty$ of the truncated spherical harmonics series expansion in a dynamic manner. Since particles have to fulfill given contact conditions, especially large particles often have many contact conditions, we are just able to create particles with nearly spherical shapes if we have "more degrees of freedom" for some particles. By increasing $L$, we can provide "more degrees of freedom" to particles with many contact conditions which is related to a large coordination number for such particles indicated by the connectivity graph $G$. Therefore, for each particle, we relate $L$ with its coordination number given by $G$. In particular, a small coordination number implies a smaller value of $L$ to guarantee sufficiently smooth particle surfaces. In contrast, a large coordination number requires a larger $L$ to avoid deformed and strongly non-spherical particles which would occur due to the many contact conditions. We allocate the value of $L$ for each particle as follows.

We set a "default", minimum and maximum degree of accuracy denoted by $L^{\scriptscriptstyle{df}}, L^{\scriptscriptstyle{min}}$ and $L^{\scriptscriptstyle{max}}$, respectively, where we used $L^{\scriptscriptstyle{df}}=8$, $L^{\scriptscriptstyle{min}}=5$ and $L^{\scriptscriptstyle{max}}=20$. Furthermore, we also define a "default" coordination number $\kappa^{\scriptscriptstyle{df}}$ as the integer closest to the mean coordination number $\kappa^{\scriptscriptstyle{exp}}$ observed in segmented tomographic image data.
%For rounding we used the rule "round half up".
Moreover, given the currently realized connectivity graph $G$, we know its minimum and maximum coordination number $\kappa^{\scriptscriptstyle{min}}$ and $\kappa^{\scriptscriptstyle{max}}$. In most cases, coordination numbers $\kappa\in\{\kappa^{\scriptscriptstyle{min}},\dots,\kappa^{\scriptscriptstyle{max}}\}$ in $G$ range from 0 to 25. The degree of accuracy for each particle with coordination number $\kappa$ is then given by \[L=L^{\scriptscriptstyle{df}}+\ind_{\{\kappa\leq\kappa^{\scriptscriptstyle{df}}\}}\frac{\kappa-\kappa^{\scriptscriptstyle{df}}}{\kappa^{\scriptscriptstyle{df}}-\kappa^{\scriptscriptstyle{min}}}(L^{\scriptscriptstyle{df}}-L^{\scriptscriptstyle{min}})+\ind_{\{\kappa>\kappa^{\scriptscriptstyle{df}}\}}\frac{\kappa-\kappa^{\scriptscriptstyle{df}}}{\kappa^{\scriptscriptstyle{max}}-\kappa^{\scriptscriptstyle{df}}}(L^{\scriptscriptstyle{max}}-L^{\scriptscriptstyle{df}}).\]
This allocation ensures a preferably uniform spread of $L$ over the range of coordination numbers $\kappa$. Furthermore, it assigns $L^{\scriptscriptstyle{df}}$ to the coordination number $\kappa^{\scriptscriptstyle{df}}$, $L^{\scriptscriptstyle{min}}$ to the coordination number $\kappa^{\scriptscriptstyle{min}}$ and $L^{\scriptscriptstyle{max}}$ to the coordination number $\kappa^{\scriptscriptstyle{max}}$.

\subsubsection{Using the whole framework to create individual particles}\label{subsubsec:particleCreation}

Now, we are in a position to place individual particles into the corresponding particle polytopes $\{P_n\}\subset\mathcal{T}$, where each particle has its origin in the centroid $M_n$ of the corresponding polytope $P_n$. For each particle we have strong control over its size through the first coefficient $a_{0,0}$. Its shape depends on the angular power function $A(l)$, the degree of accuracy $L$ and the points in $\mathcal{C}_n$ which the particle has to cross, where the points in $\mathcal{C}_n$ are directly influenced by the shape of $P_n$. Recall that the contact conditions given by the points in $\mathcal{C}_n$ ensure connectedness of a particle as indicated by the graph $G$ from Section~\ref{subsec:graphModel}. They are translated into terms for the spherical harmonics series expansion of a particle in the following way. Each of the $N$ points $C_i\in\mathcal{C}_n$ can be expressed in spherical coordinates with respect to the centroid $M_n$ of polytope $P_n$ into which we will place a particle. This means that a point $C_i$ is given as $(\theta_i, \phi_i, r_i)$ for $i\in\{1,\dots,N\}$, where the angles $\theta_i$ and $\phi_i$ describe the direction (of the vector) from $M_n$ to $C_i$ and $r_i=\|M_n-C_i\|$ is the distance between $M_n$ and $C_i$. Then, for a particle being placed into $P_n$, it has to hold that \[\psi(\theta_1,\phi_1)=r_1,\dots,\psi(\theta_N,\phi_N)=r_N,\] where each $\psi(\theta_i,\phi_i)=r_i$ is a linear equation with the (random) unknowns $\allowbreak a_{1,0},\allowbreak \mathrm{Re}(a_{1,1}),\allowbreak \mathrm{Im}(a_{1,1}),\allowbreak\dots,\allowbreak \mathrm{Re}(a_{L,L}),\allowbreak \mathrm{Im}(a_{L,L})$. Since $a_{0,0}$ is a constant which is known from Section~\ref{subsubsec:aps}, we can rewrite the system of linear equations as \[\hat{\psi}(\theta_i,\phi_i)=r_i-a_{0,0}Y_{0,0}(\theta_i,\phi_i)\quad\text{for } i=1,\dots,N,\] where $\hat{\psi}(\theta_i,\phi_i)$ corresponds to $\psi(\theta_i,\phi_i)$ but without the term $a_{0,0}Y_{0,0}(\theta_i,\phi_i)$. The particle is then created by drawing a realization from the normal distributed coefficient vector $(a_{1,0},\allowbreak \mathrm{Re}(a_{1,1}),\allowbreak \mathrm{Im}(a_{1,1}),\allowbreak\dots,\allowbreak \mathrm{Re}(a_{L,L}),\allowbreak \mathrm{Im}(a_{L,L}))^{\mathsf T}$ given $\hat{\psi}(\theta_1,\phi_1)=r_1-a_{0,0}Y_{0,0}(\theta_1,\phi_1),\allowbreak\dots,\allowbreak\hat{\psi}(\theta_N,\phi_N)=r_N-a_{0,0}Y_{0,0}(\theta_N,\phi_N)$. The detailed procedure how such a conditionally normal distributed vector is simulated can be found in~\cite{feinauer.2015b}. Finally, the centroid $M_n$ and the coefficient vector $(a_{0,0},\allowbreak a_{1,0},\allowbreak \mathrm{Re}(a_{1,1}),\allowbreak \mathrm{Im}(a_{1,1}),\allowbreak\dots,\allowbreak \mathrm{Re}(a_{L,L}),\allowbreak \mathrm{Im}(a_{L,L}))^{\mathsf T}$ describe the realized particle being placed into polytope $P_n$.

Sometimes we have to struggle with the case that an undesired and degenerated particle is created. On the one hand, a particle is identified as degenerated in the same way as in~\cite{westhoff.2017}, namely, if the maximum particle extension $\max\{\psi(\theta,\phi)\}$ is larger than 1.5 times the maximum distance of the centroid $M_n$ to the boundary of its polytope. Furthermore, a particle is deemed to be degenerated if it has a negative extension, i.e., if $\psi(\theta,\phi)<0$ for some direction $(\theta,\phi)$, which is possible by definition. In the case that a particle is identified as degenerated, we first try to handle this problem by running up to 1000 repetitions to create a new particle, i.e., drawing new realizations from a conditionally normal distributed random vector. If this still results in a degenerated particle (which occurs in very few, if any, instances), then we ultimately create a particle with coefficient vector $(\tilde{a}_{0,0},\allowbreak 0,\allowbreak\dots,0)^{\mathsf T}$, where $\tilde{a}_{0,0}=r_{part}/Y_{0,0}(\theta,\phi)=2\sqrt{\pi}r_{part}$. Such a particle is simply a sphere with center $M_n$ and radius $r_{part}$. For $r_{part}$ we distinguish between two situations. If the connectivity graph indicates at least one connection for the particle, then $r_{part}$ is the maximum distance of the polytope centroid to a point of the corresponding contact conditions, i.e., $r_{part}=\max_{i\in\{1,\dots,N\}}\{r_i\}$. A sphere with this radius preserves all contact conditions. If the connectivity graph does not indicate a connection for the particle, then $r_{part}$ is equal to the mark of the marked point which induced the corresponding particle polytope, i.e., $r_{part}=R_n$, where $(S_n,R_n)\in\mathcal{S}$ induced polytope $P_n$. A sphere with such a radius mimics a particle of the originally target size. Note that the idea to ultimately describe a particle through a sphere is reasonable, since the particles in the considered cathode materials exhibit nearly spherical shapes.

\subsubsection{Morphological smoothing of the particle system}\label{subsubsec:morphSmooth}

The stochastic 3D microstructure modeling is completed by the following step. First, we discard all auxiliary tools such as the marked point patterns, the Laguerre tessellation or the connectivity graph and only keep the created particle system which might have a partially rough surface resulting in a too large surface area. To avoid this, we perform a morphological closing of the particle phase with a ball of radius 2 voxels as structuring element. Subsequently, we proceed with an opening of the particle phase with the same structuring element. For more information about morphological closing and opening we refer, e.g., to~\cite{beucher.1993} and~\cite{soille.2003}.

\section{Model validation}\label{sec:valid}

After having introduced the stochastic 3D modeling approach for micro\-structures of cathodes, we perform a validation by comparing several image characteristics which are computed, on the one hand, from (segmented) tomographic image data and, on the other hand, from realizations of the model. Thereby, we will demonstrate that the model achieves good fits for all considered image characteristics regardless of the scenario \textsc{P}, \textsc{A1} or \textsc{A2}. All characteristics are computed on images with a size of $400\times 400$ voxels in horizontal direction and either $z_{\textsc{P}}$, $z_{\textsc{A1}}$ or $z_{\textsc{A2}}$ voxels in vertical direction depending on the given scenario. That is, for each tomographic data set, we cut out four disjoint subimages with the corresponding size from the larger image (which has $1000\times 1000$ voxels in horizontal direction, see Section~\ref{subsec:data_prepro}), since four is the maximum number of disjoint cutouts having a size of $400\times 400$ voxels that can be extracted from it. On the other hand, four realizations of the model are generated in windows with the same size.

In the following, the depicted result of each considered characteristic is the mean over the results of this characteristic computed on four tomographic cutouts or four model realizations. Additionally, some of the following figures contain the ranges between the (pointwise estimated) minimum and maximum values as shaded areas, see, e.g., Figure~\ref{fig:cpsd}.

\subsection{Characteristics of pore phase}\label{subsec:porePhaseValid}

\begin{figure}[!th]
	\centering
	\begin{subfigure}[c]{0.32\textwidth}
		\includegraphics[width=\textwidth]{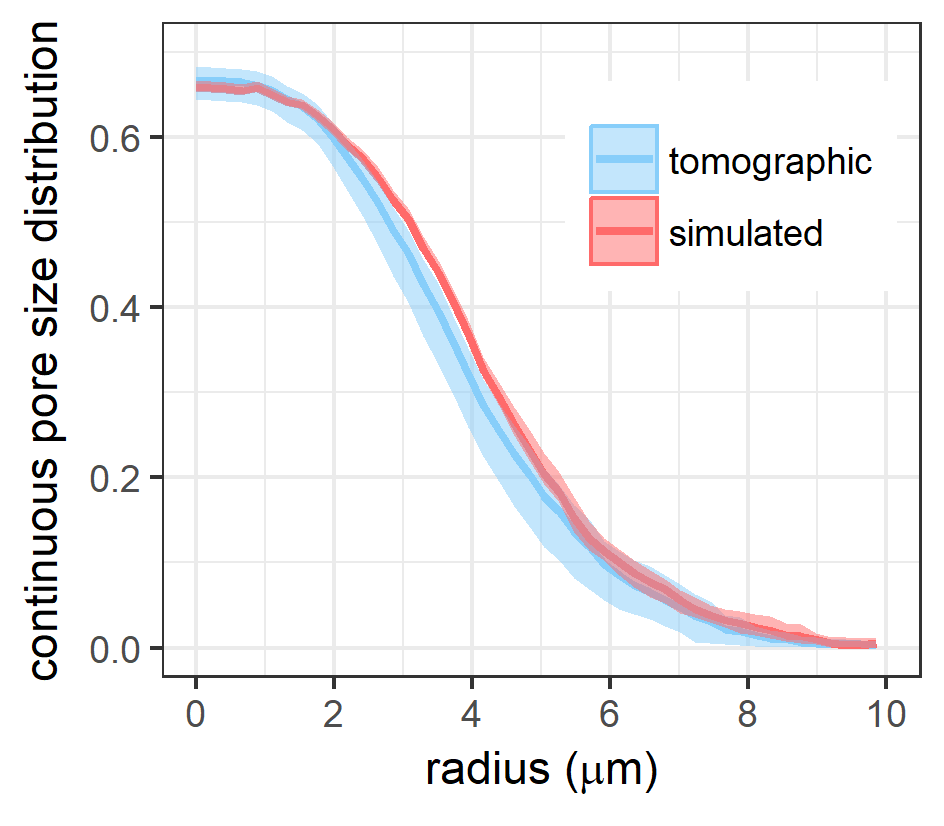}
		\caption{Scenario \textsc{P}}
		\label{subfig:cpsdP}
	\end{subfigure}
	\begin{subfigure}[c]{0.32\textwidth}
		\includegraphics[width=\textwidth]{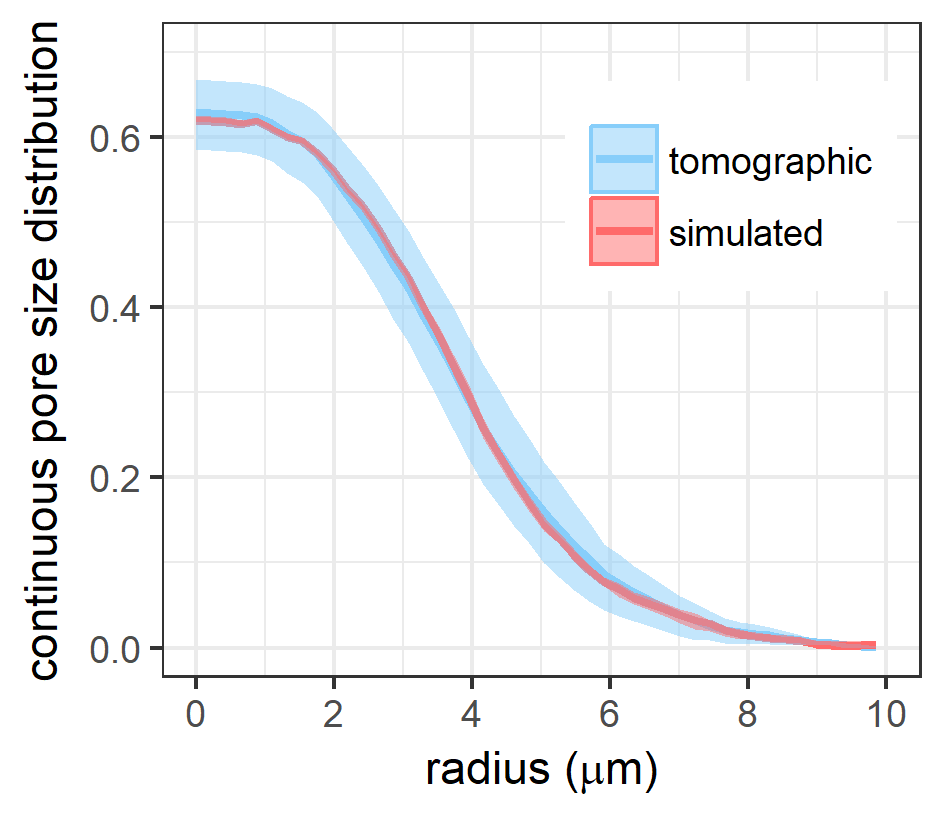}
		\caption{Scenario \textsc{A1}}
		\label{subfig:cpsdA1}
	\end{subfigure}
	\begin{subfigure}[c]{0.32\textwidth}
		\includegraphics[width=\textwidth]{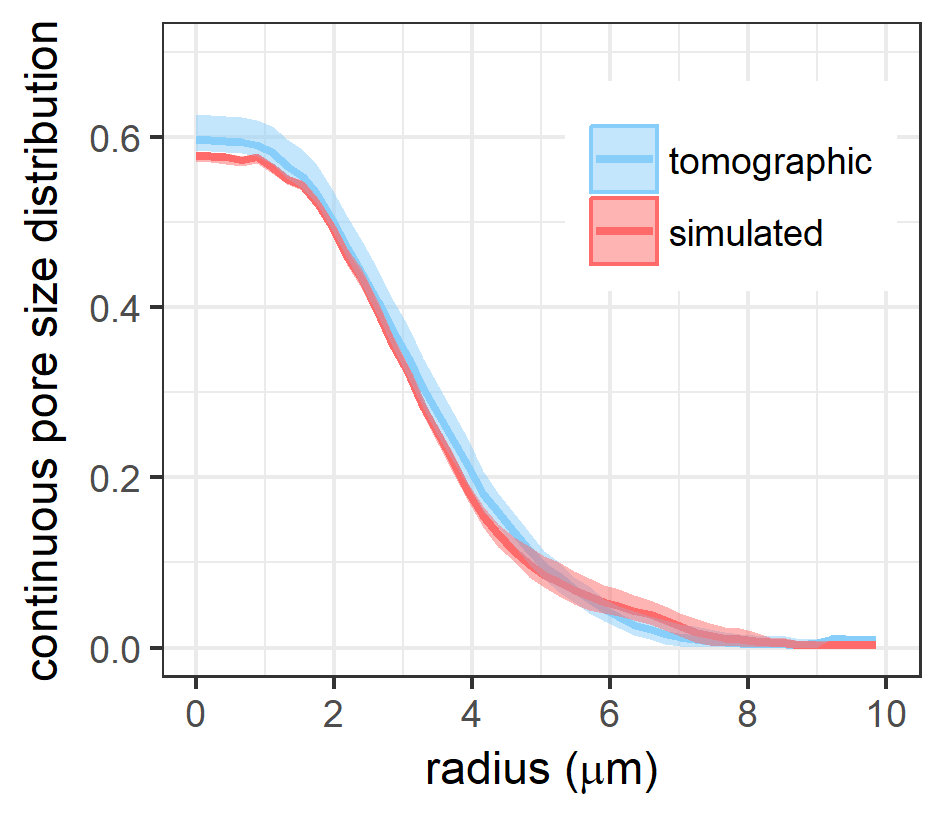}
		\caption{Scenario \textsc{A2}}
		\label{subfig:cpsdA2}
	\end{subfigure}
	\caption{Continuous pore size distribution, where the value for a radius of 0 corresponds to the overall porosity.}
	\label{fig:cpsd}
\end{figure}

We first validate the model by comparing some characteristics which show that the model is able to generate pore morphologies as observed in tomographic data sets. Recall that one of the features in the considered cathodes are the locally occurring large pores. To deal with this observation, a marked point pattern was introduced which is supposed to model large pores (see Section~\ref{subsec:pointModel}). Furthermore, the additional pore polytopes from Section~\ref{subsubsec:porePoly} also contribute to a suitable pore morphology. A characteristic perfectly depicting that the model leads to suitable pore morphologies is the so-called continuous pore size distribution (see~\cite{muench.2008} for details). It describes how much volume of the pore phase can be covered by (overlapping) spheres of a fixed radius or, in other words, how much porosity can be accessed if we fill in the pore phase with such spheres. A good accordance, especially for pores with large radii, between the continuous pore size distributions for tomographic and simulated data holds for all three scenarios \textsc{P}, \textsc{A1}, \textsc{A2} and is depicted in Figure~\ref{fig:cpsd}.

\begin{figure}[!th]
	\centering
	\begin{subfigure}[c]{0.32\textwidth}
		\includegraphics[width=\textwidth]{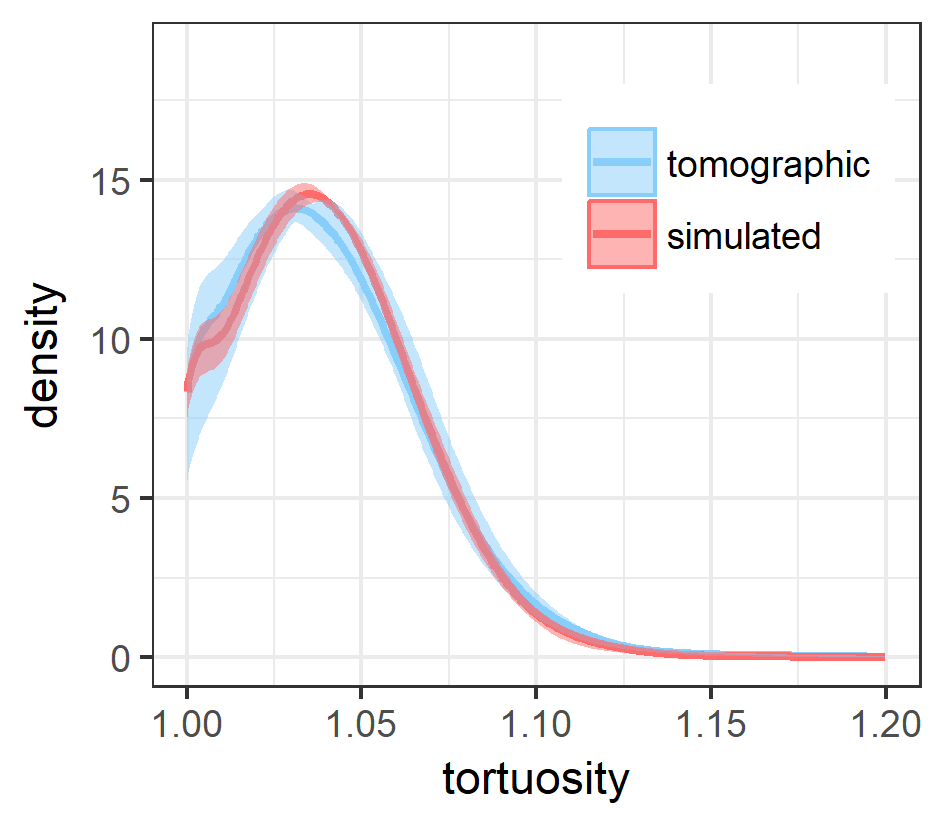}
		\caption{Scenario \textsc{P}}
		\label{subfig:geodTortP}
	\end{subfigure}
	\begin{subfigure}[c]{0.32\textwidth}
		\includegraphics[width=\textwidth]{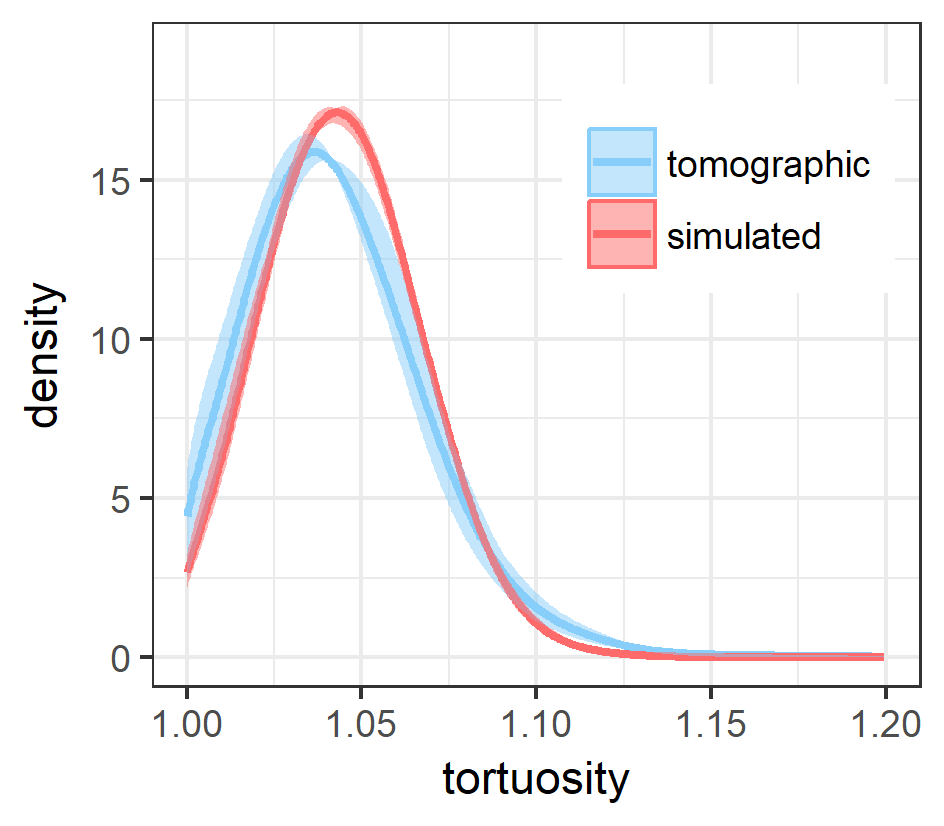}
		\caption{Scenario \textsc{A1}}
		\label{subfig:geodTortA1}
	\end{subfigure}
	\begin{subfigure}[c]{0.32\textwidth}
		\includegraphics[width=\textwidth]{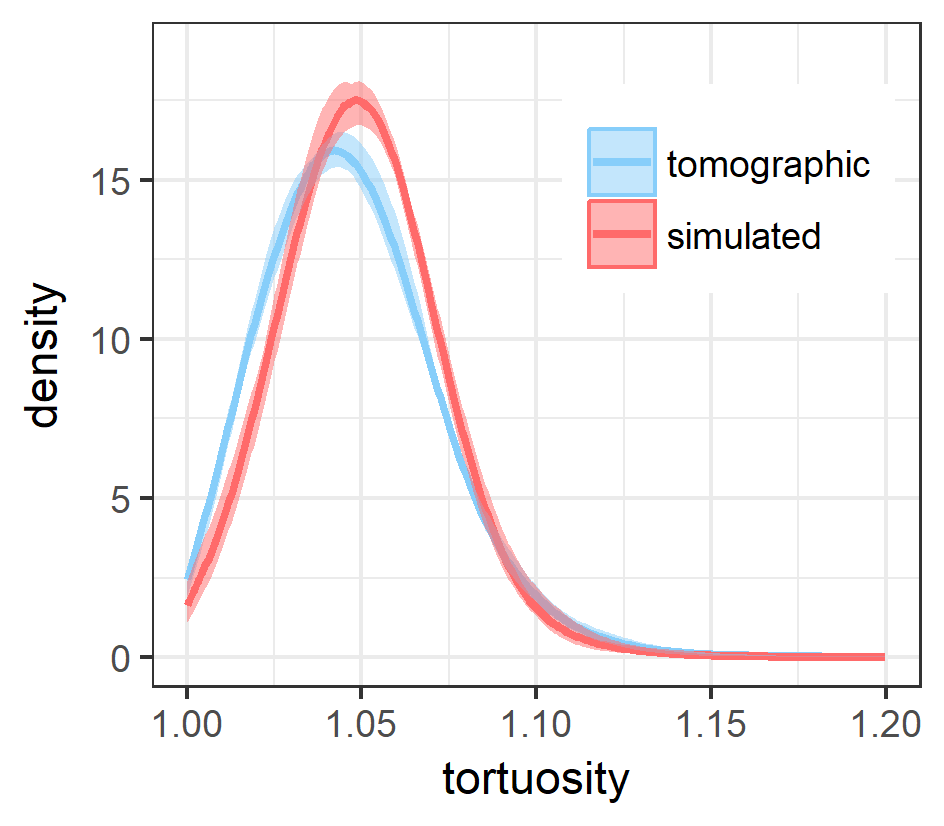}
		\caption{Scenario \textsc{A2}}
		\label{subfig:geodTortA2}
	\end{subfigure}
	\caption{Geodesic tortuosity of paths through pore phase in through-plane direction.}
	\label{fig:geodTort}
\end{figure}

The morphology of pore phase plays an important role regarding transport properties of cathodes, since lithium ions flow through the pore phase during charging and discharging of the batteries. In particular, an appropriate characteristic to evaluate the geometry of transport paths through the pore phase is its tortuosity and, more precisely, the so-called geodesic tortuosity~\cite{stenzel.2016}, which describes the ratio of the length of shortest paths through the pore phase in through-plane direction divided by the thickness of the cathode. Figure~\ref{fig:geodTort} shows the distribution of these ratios for randomly chosen starting points on the "bottom" of the cathode in through-plane direction.%Note that we cannot see any significant differences when starting from one or the other surface of the cathode in through-plane direction.

%\textcolor{blue}{possible characteristics: cpsd, (mean) tortuosities, background points (marks, intensity), surfaceBackCells, volumeBackCells, areaMatchingMainBack (sehe darin keinen relevanten Informationsgehalt),...}

\subsection{Characteristics of particle phase}\label{subsec:particlePhaseValid}

\begin{figure}[!ht]
	\centering
	\begin{subfigure}[t]{0.49\textwidth}
		\includegraphics[width=\textwidth]{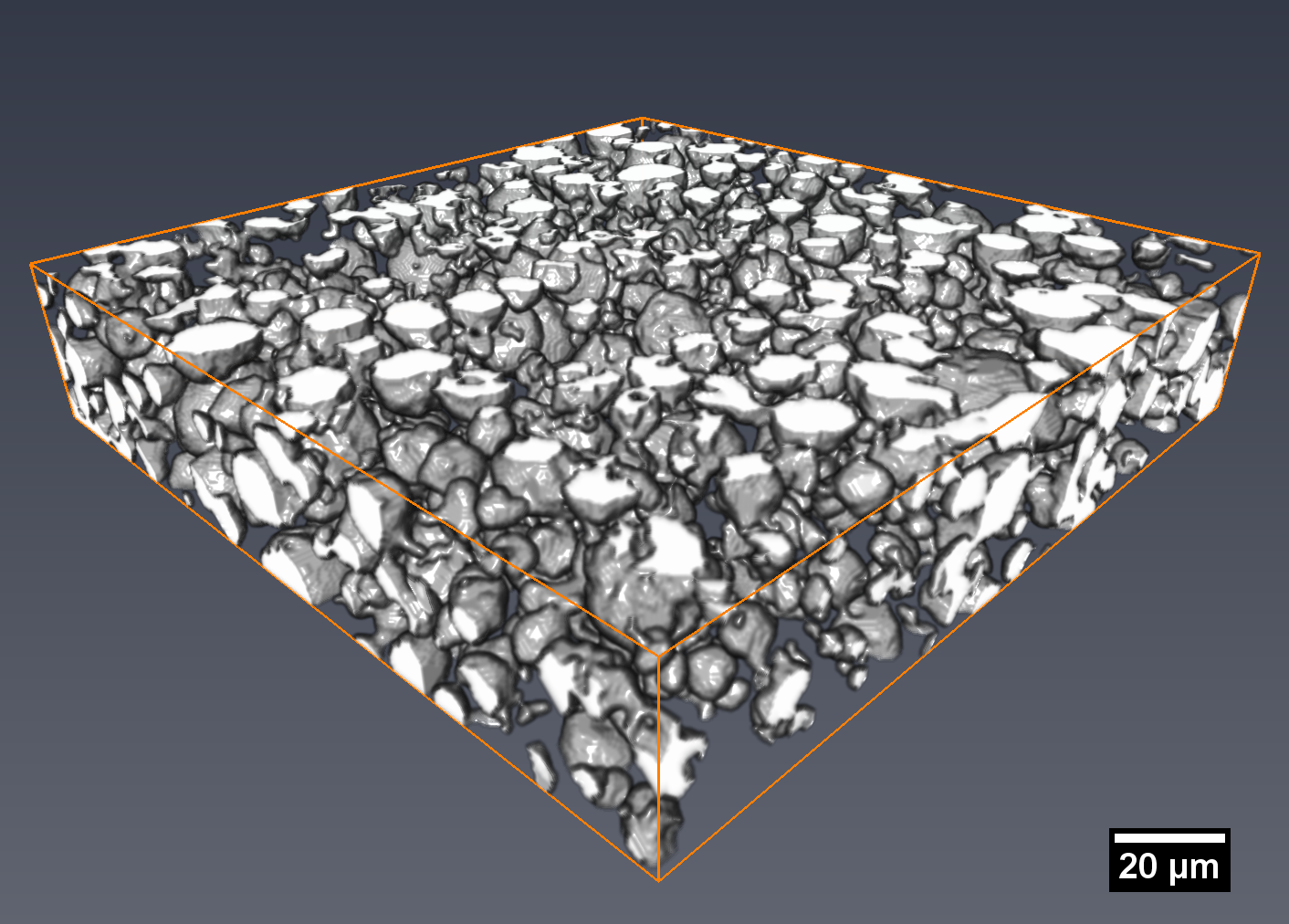}
		\caption{Tomographic image data}
		\label{subfig:tomoCutout}
	\end{subfigure}
	\begin{subfigure}[t]{0.49\textwidth}
		\includegraphics[width=\textwidth]{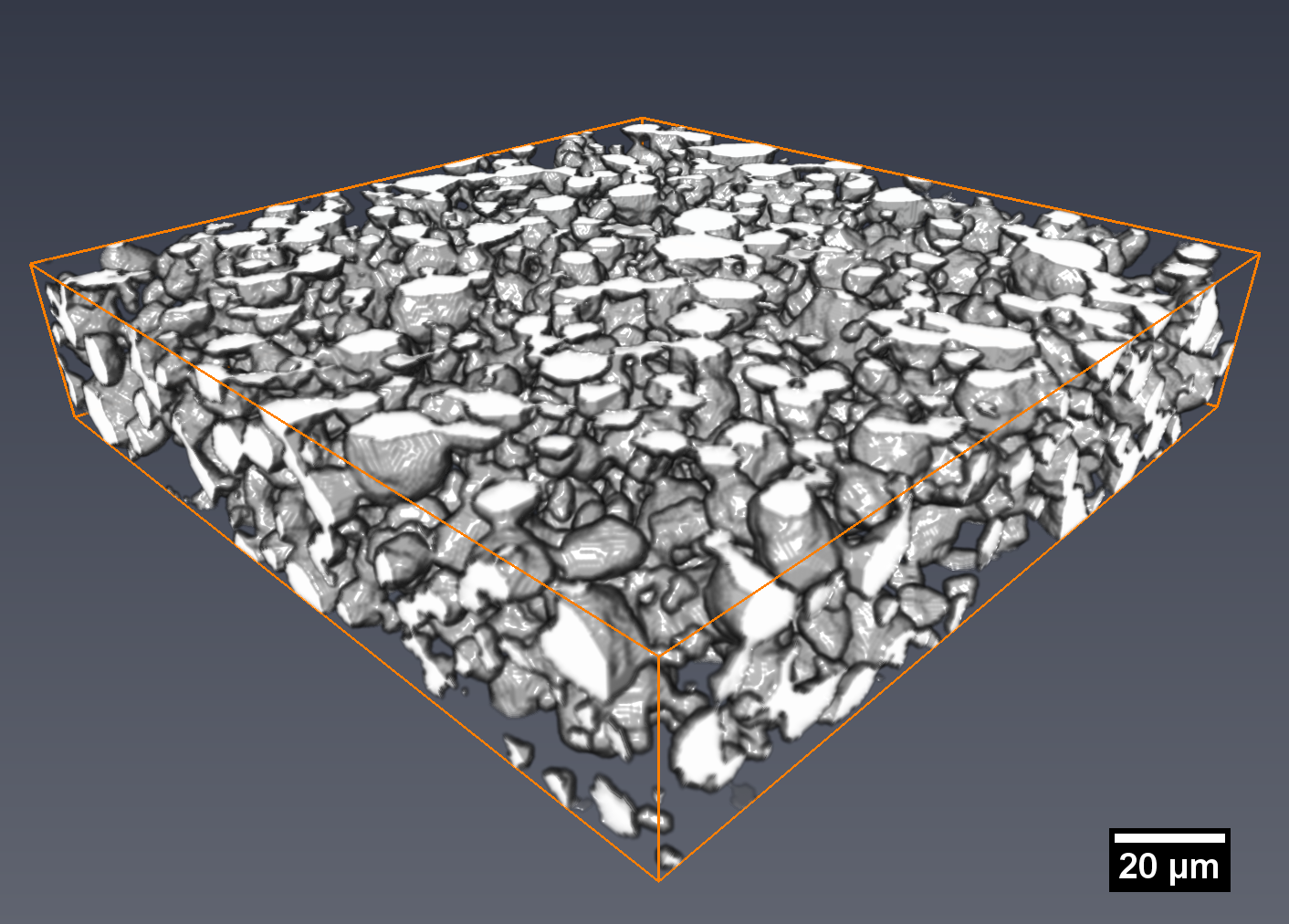}
		\caption{Simulated image data}
		\label{subfig:simCutout}
	\end{subfigure}
	\caption{3D renderings of microstructure cutouts of the pristine cathode (scenario \textsc{P}).}
	\label{fig:cutoutComp3D}
\end{figure}

We now validate the model by comparing the particle systems of tomographic and simulated image data. Visual inspection of the 3D cutouts described at the beginning of Section~\ref{sec:valid} shows a very good accordance for the pristine cathode (scenario \textsc{P}), see Figure~\ref{fig:cutoutComp3D}, which also holds for the other two scenarios. We can recognize a suitable pore morphology and, in particular, the particles seem to have appropriate locations, sizes and shapes. To confirm this, we take a closer look at several characteristics of the particle phase.

To begin with, we compare so-called first order characteristics, namely, volume fraction and specific surface area (i.e., surface area divided by bulk volume) of the particle phase. Table~\ref{tab:volSSA} shows the mean values and relative errors listed for scenarios \textsc{P}, \textsc{A1} and \textsc{A2}. %, whereas Figure~\ref{fig:boxplotsVolSSA} depicts boxplots for the pristine cathode.
The volume fraction is well controlled by the model (see also Section~\ref{subsubsec:aps}), just the values of specific surface area are a bit too large for simulated data of scenarios \textsc{A1} and \textsc{A2}.

\begin{table}[!th]
	\centering
	\begin{tabular}{r|c|c|c|c|c|c}
		\hline
		& \multicolumn{3}{c|}{volume fraction} & \multicolumn{3}{|c}{specific surface area ($\unit[1/]{\mu m}$)}\\
		& \textsc{P} & \textsc{A1} & \textsc{A2} & \textsc{P} & \textsc{A1} & \textsc{A2}\\
		\hline
		tomographic & 0.3358 & 0.3737 & 0.4144 & 0.1958 & 0.1991 & 0.2199\\
		simulated & 0.3342 & 0.3758 & 0.4131 & 0.1945 & 0.2087 & 0.2353\\
		relative error & 0.48\% & 0.56\% & 0.31\% & 0.66\% & 4.8\% & 6.98\%\\
		\hline
	\end{tabular}
	\caption{Mean value and relative error of volume fraction and specific surface area of the particle phase.}
	\label{tab:volSSA}
\end{table}
%\begin{figure}[!ht]
	%\centering
	%\begin{subfigure}[t]{0.4\textwidth}
		%\includegraphics[width=\textwidth]{images/volumeP.png}
		%\caption{Volume fraction}
		%\label{subfig:boxplotVol}
	%\end{subfigure}
	%\begin{subfigure}[t]{0.4\textwidth}
		%\includegraphics[width=\textwidth]{images/surfaceP.png}
		%\caption{Specific surface area}
		%\label{subfig:boxplotSSA}
	%\end{subfigure}
	%\caption{\textcolor{red}{TODO: etwas anderes alles boxplots da nur 4 Datenpunkte} Boxplots (boxes show range between 25\% and 75\% quantiles) for volume fraction and specific surface area of the particle phase, considered for the pristine cathode (scenario \textsc{P}).}
	%\label{fig:boxplotsVolSSA}
%\end{figure}

\begin{figure}[!th]
	\centering
	\begin{subfigure}[c]{0.32\textwidth}
		\includegraphics[width=\textwidth]{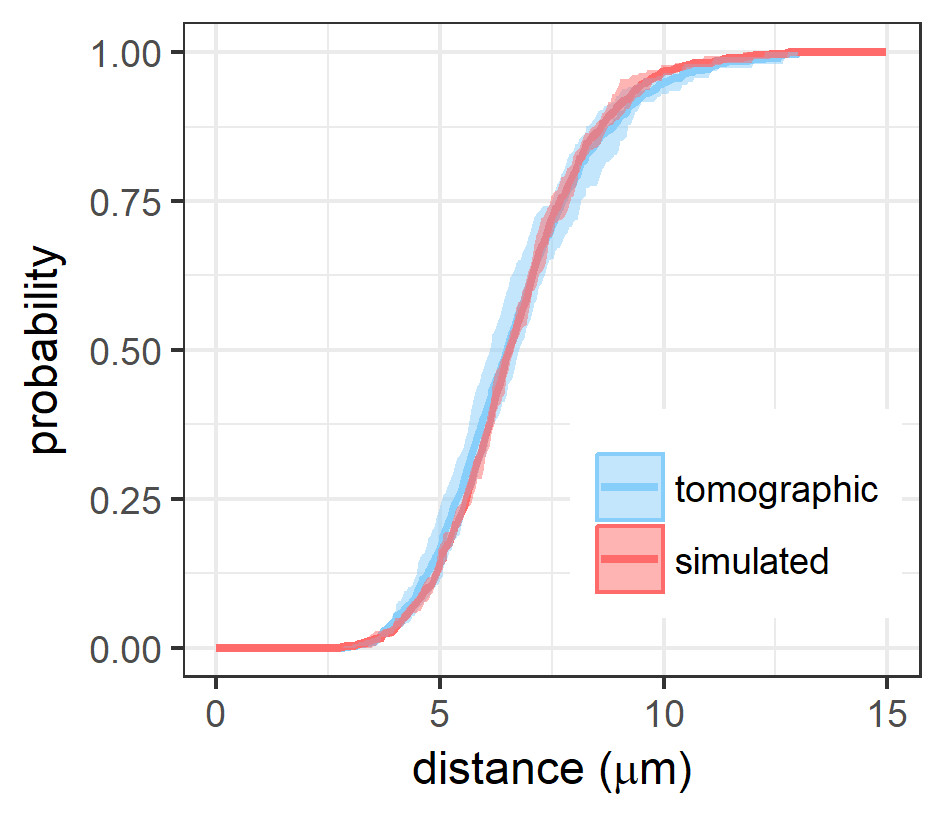}
		\caption{Scenario \textsc{P}}
		\label{subfig:nnP}
	\end{subfigure}
	\begin{subfigure}[c]{0.32\textwidth}
		\includegraphics[width=\textwidth]{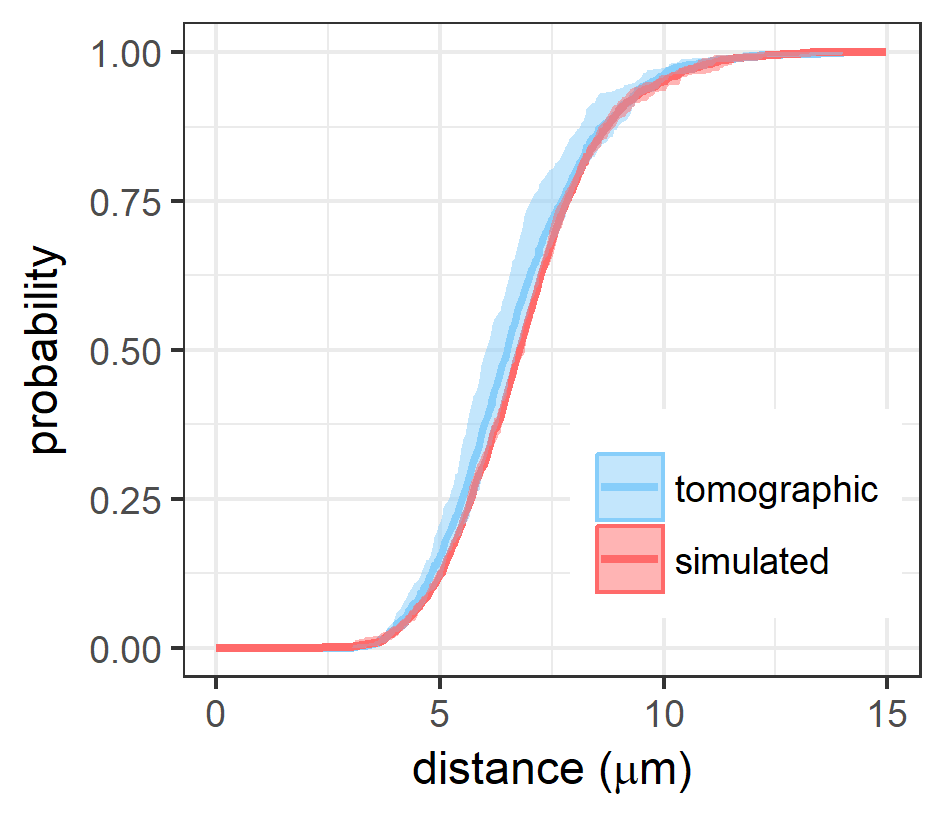}
		\caption{Scenario \textsc{A1}}
		\label{subfig:nnA1}
	\end{subfigure}
	\begin{subfigure}[c]{0.32\textwidth}
		\includegraphics[width=\textwidth]{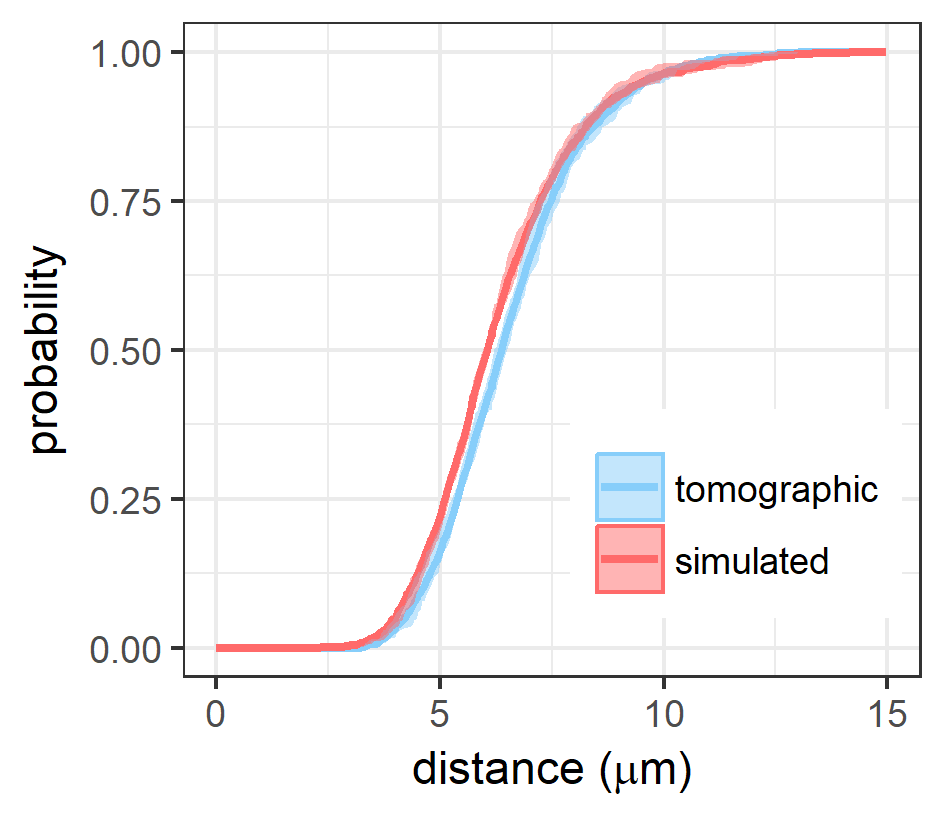}
		\caption{Scenario \textsc{A2}}
		\label{subfig:nnA2}
	\end{subfigure}
	\caption{Nearest neighbor distance distribution function.}
	\label{fig:nn}
\end{figure}

\begin{figure}[!t]
	\centering
	\begin{subfigure}[c]{0.32\textwidth}
		\includegraphics[width=\textwidth]{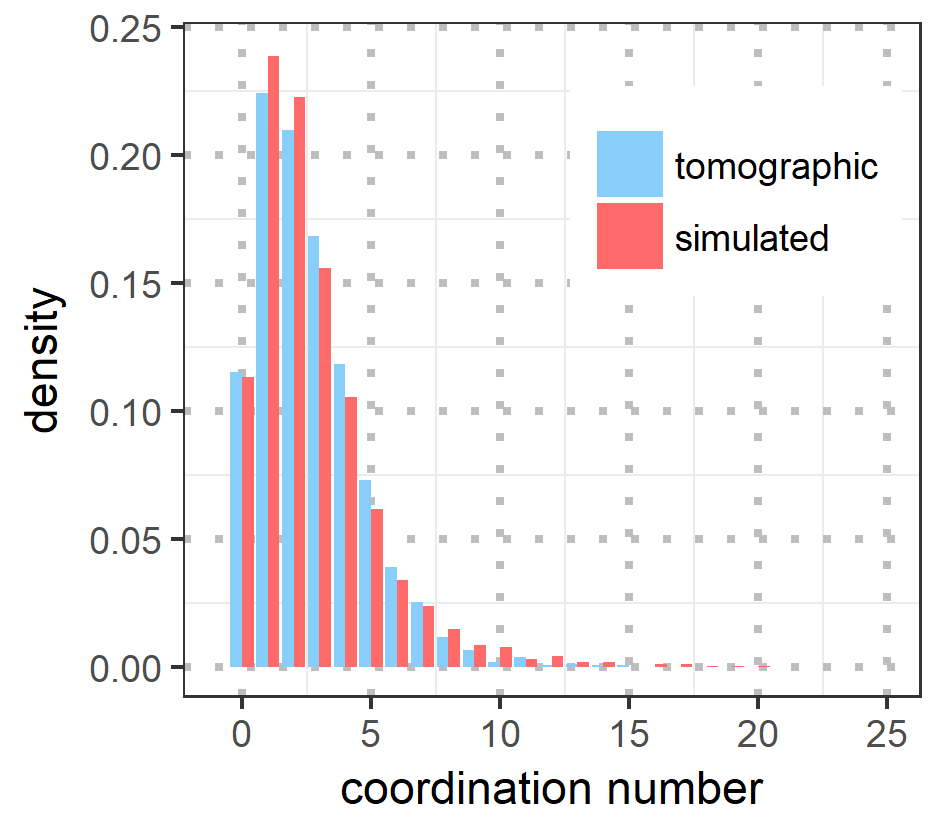}
		\caption{Scenario \textsc{P}}
		\label{subfig:coordNumP}
	\end{subfigure}
	\begin{subfigure}[c]{0.32\textwidth}
		\includegraphics[width=\textwidth]{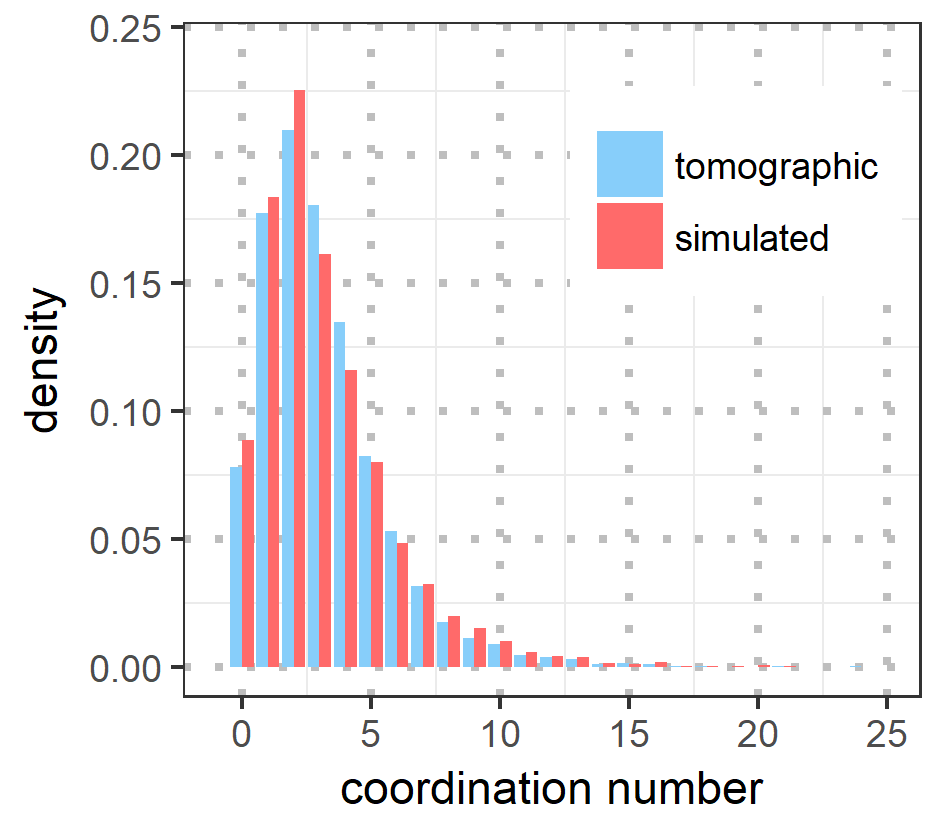}
		\caption{Scenario \textsc{A1}}
		\label{subfig:coordNumA1}
	\end{subfigure}
	\begin{subfigure}[c]{0.32\textwidth}
		\includegraphics[width=\textwidth]{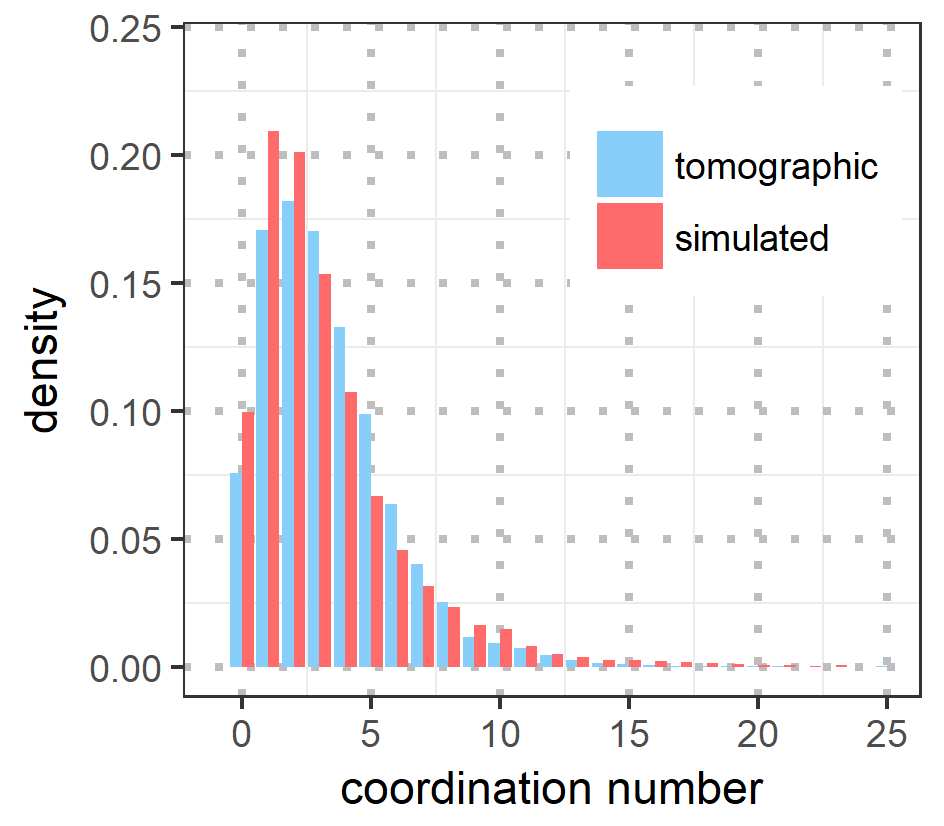}
		\caption{Scenario \textsc{A2}}
		\label{subfig:coordNumA2}
	\end{subfigure}
	\caption{Histograms of the coordination numbers, where the first two bins (at 0) indicate the relative frequencies of unconnected particles.}
	\label{fig:coordNum}
\end{figure}

\begin{figure}[!th]
	\centering
	\begin{subfigure}[c]{0.32\textwidth}
		\includegraphics[width=\textwidth]{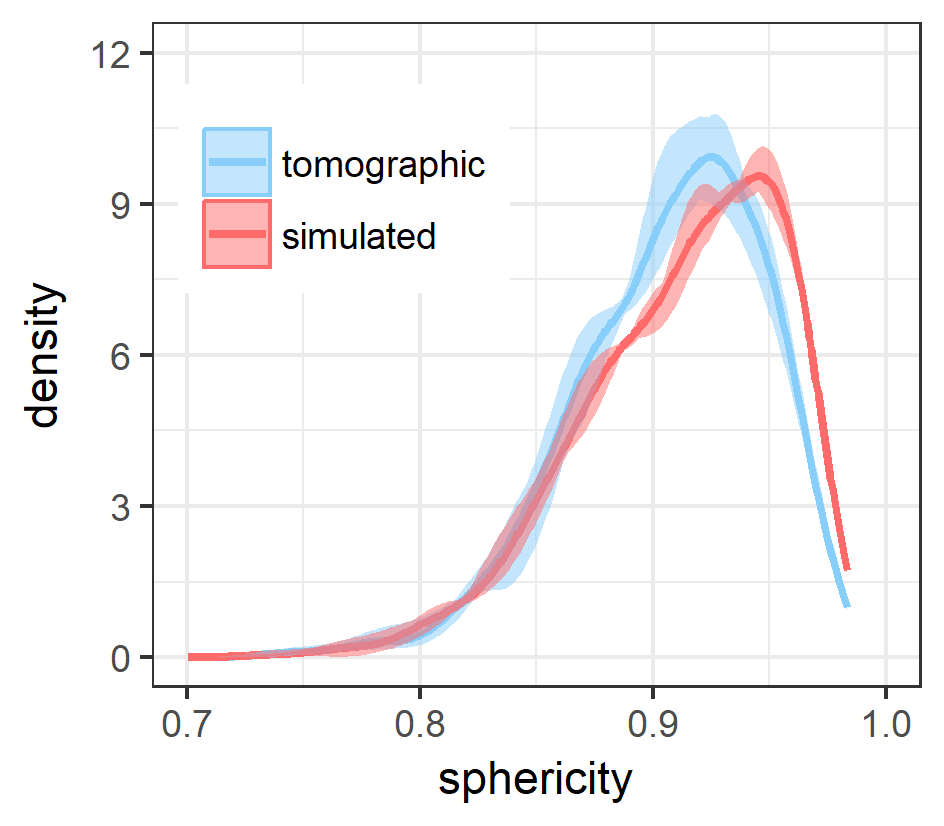}
		\caption{Scenario \textsc{P}}
		\label{subfig:sphericityPolyP}
	\end{subfigure}
	\begin{subfigure}[c]{0.32\textwidth}
		\includegraphics[width=\textwidth]{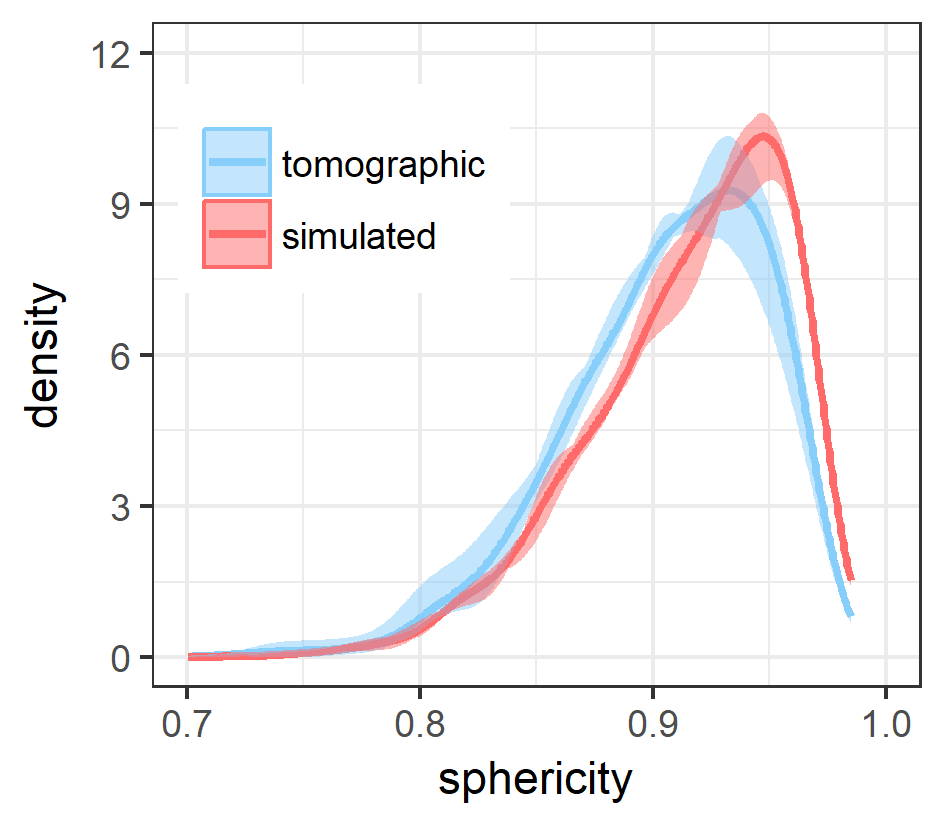}
		\caption{Scenario \textsc{A1}}
		\label{subfig:sphericityPolyA1}
	\end{subfigure}
	\begin{subfigure}[c]{0.32\textwidth}
		\includegraphics[width=\textwidth]{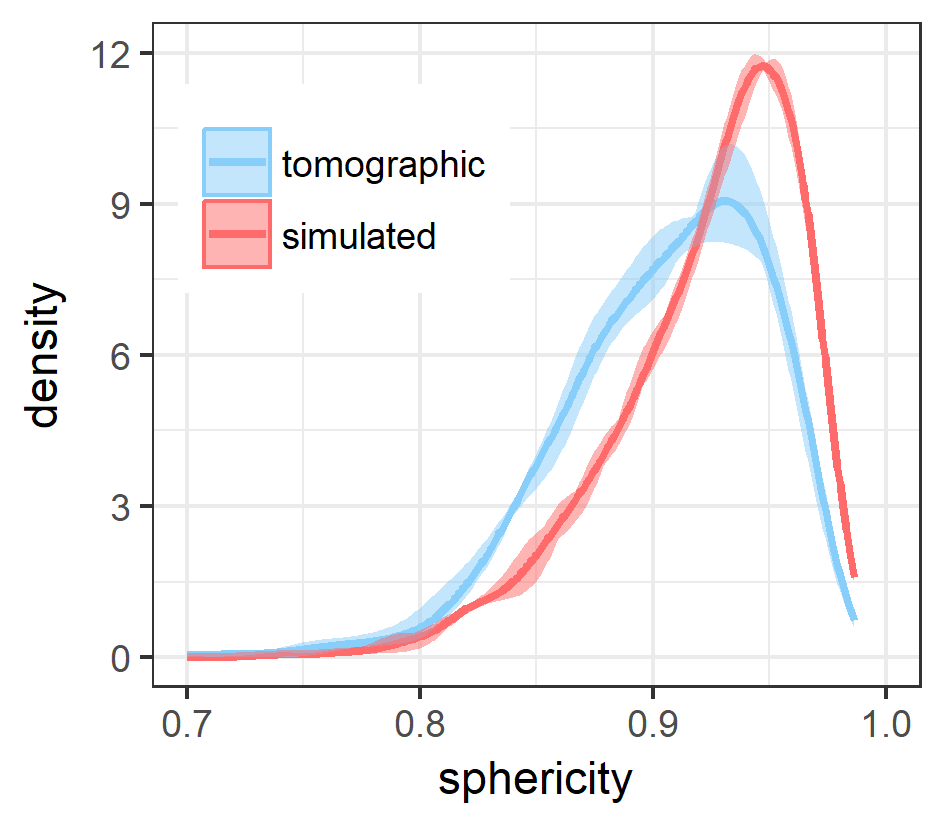}
		\caption{Scenario \textsc{A2}}
		\label{subfig:sphericityPolyA2}
	\end{subfigure}
	\caption{Distribution of sphericity of particle polytopes.}
	\label{fig:sphericityPoly}
\end{figure}

Before we continue to consider characteristics which evaluate the particle phase in its entirety, we first look at some characteristics which help us to validate single components of the model. These are the models of marked point patterns, connectivity graph and Laguerre tessellation, where in Section~\ref{subsec:pointModel} the final marked point pattern $\mathcal{S}$ has described the approximate locations and sizes of particles. Note that we do not have to give thought to the intensity of these points (i.e., the expected number of particles per unit volume) and the distribution of their marks, because we directly adjust these two model components to the features which we observe in tomographic data. Thus, we are just interested in the structural arrangement of these points and compare it to the arrangement of particle midpoints in tomographic data. The so-called nearest neighbor distance distribution~\cite{illian.2008} is a useful characteristic to evaluate the spatial arrangement of points. It describes the probability of a randomly chosen point of the point pattern to find its nearest neighbor within some given distance. The corresponding probability distribution functions for a given range of distances are shown in Figure~\ref{fig:nn} and exhibit an excellent fit.
Next, we briefly check whether the simulated connectivity graph from Section~\ref{subsec:graphModel} mimics the graph which describes particle connectivity in (segmented) tomographic data. For this purpose, we consider the distribution of the coordination number, i.e., the number of directly connected particles per particle. Histograms of the coordination numbers are shown in Figure~\ref{fig:coordNum}. The fits of the distributions are very good and, of course, also the mean coordination numbers are correct since they have been adjusted when we determined the correction factor $c$ at the end of Section~\ref{subsubsec:caliConnGraphModel}. Furthermore, the relative frequencies of particles having no connection are nicely fitted by the graph model, which is indicated by the first two bins (at 0) of the histograms in Figure~\ref{fig:coordNum}. 
To obtain reasonably shaped particles, especially nearly spherical particles, the shapes of the convex polytopes for particles $\{P_n\}$ of the final Laguerre tessellation $\mathcal{T}$ introduced in Section~\ref{subsubsec:porePoly} are very important. In particular, the enhanced insertion of further pore polytopes in Section~\ref{subsubsec:porePoly} aims at polytopes for particles having, among others, nearly spherical shapes. Then, it is also more likely that the particles under contact conditions have nearly spherical shapes. As an indicator of spherical-shaped particle polytopes we consider the distribution of sphericity of these polytopes, where sphericity means how spherical an object is compared to an ideal sphere. More precisely, the sphericity $\Psi$ is equal to $\pi^{\frac{1}{3}}(6V_{obj})^{\frac{2}{3}}/A_{obj}$, where $V_{obj}$ is the volume and $A_{obj}$ is the surface area of the considered object (e.g., a polytope), see~\cite{wadell.1935}. The corresponding distributions are depicted in Figure~\ref{fig:sphericityPoly}, where it is clearly visible that the simulated results correspond to the tomographic observations very well.

Finally, we return to characteristics which evaluate the entire morphology of the particle phase. A characteristic giving us an insight into the 3D morphology of the particle phase is the so-called chord length distribution~\cite{ohser.2000}. Let $F_{\omega}\colon[0,\infty)\to[0,1]$ denote the probability distribution function of the length of segments (chords) which run through the particle phase in a predefined direction $\omega$, e.g., in through-plane direction. In other words, if a randomly chosen straight line in direction $\omega$ crosses the cathode, then segments of intersection with the particle phase occur. The probability that the length of such a (random) segment is less than or equal to $c\geq 0$ is then given by $F_{\omega}(c)$. In this paper, we analyze the chord length distribution in through-plane and two perpendicular in-plane directions.
\begin{figure}[!ht]
	\centering
	\begin{subfigure}[c]{0.4\textwidth}
		\includegraphics[width=\textwidth]{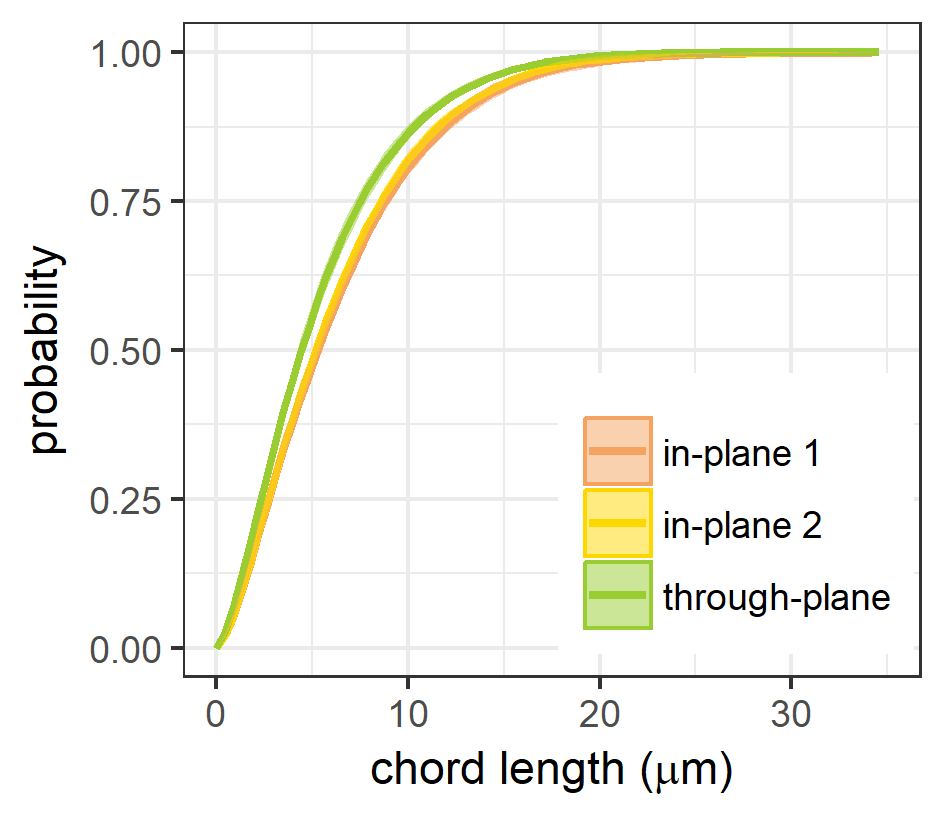}
		\caption{Tomographic}
		\label{subfig:chordTomoP}
	\end{subfigure}
	\begin{subfigure}[c]{0.4\textwidth}
		\includegraphics[width=\textwidth]{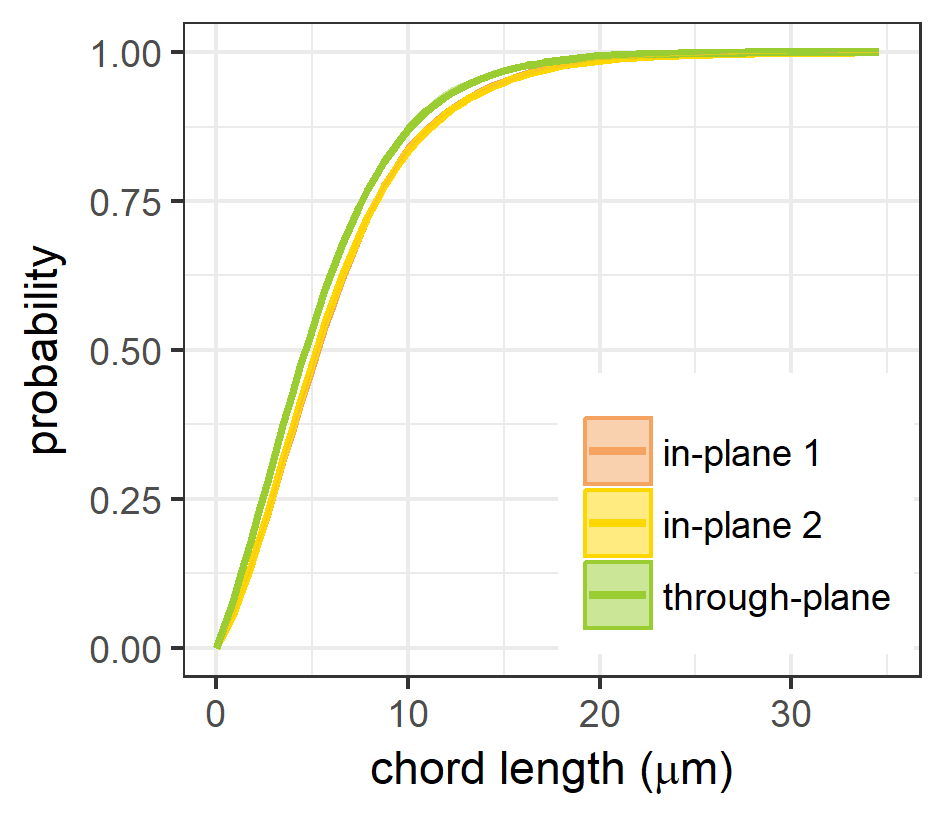}
		\caption{Simulated}
		\label{subfig:chordSimP}
	\end{subfigure}
	\caption{Chord length distributions for the pristine cathode from tomographic and simulated data. In case of tomographic data, the slightly left\hyp{}shifted chord length distribution in through-plane direction is just an artifact due to small thickness of the cathode and it does not result from anisotropy of the material, since there is no anisotropy in any of the three scenarios. In general, the chord lengths in simulated data are slightly smaller, but the model perfectly hits the isotropy of the cathodes.}
	\label{fig:chord}
\end{figure}
In Figure~\ref{fig:chord} we show the estimated curves for the pristine cathode (scenario \textsc{P}). The fit between the chord length distributions extracted from tomographic and simulated microstructures is also pretty good for the other two scenarios.

\begin{figure}[!ht]
	\centering
	\begin{subfigure}[c]{0.32\textwidth}
		\includegraphics[width=\textwidth]{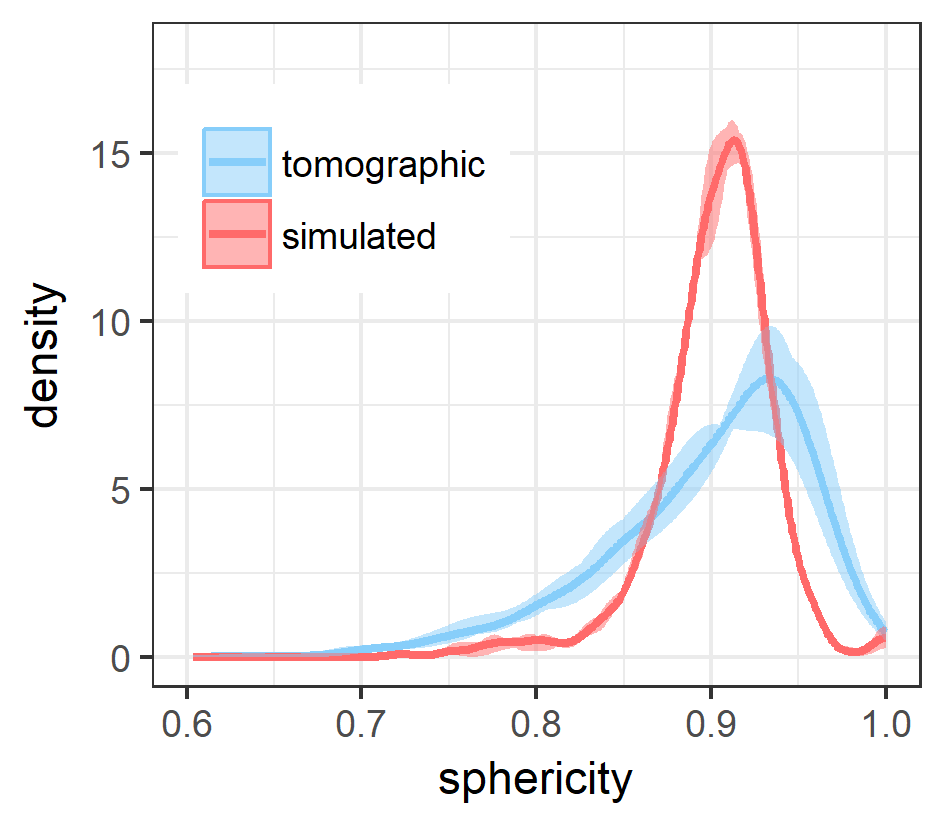}
		\caption{Scenario \textsc{P}}
		\label{subfig:sphericityPartP}
	\end{subfigure}
	\begin{subfigure}[c]{0.32\textwidth}
		\includegraphics[width=\textwidth]{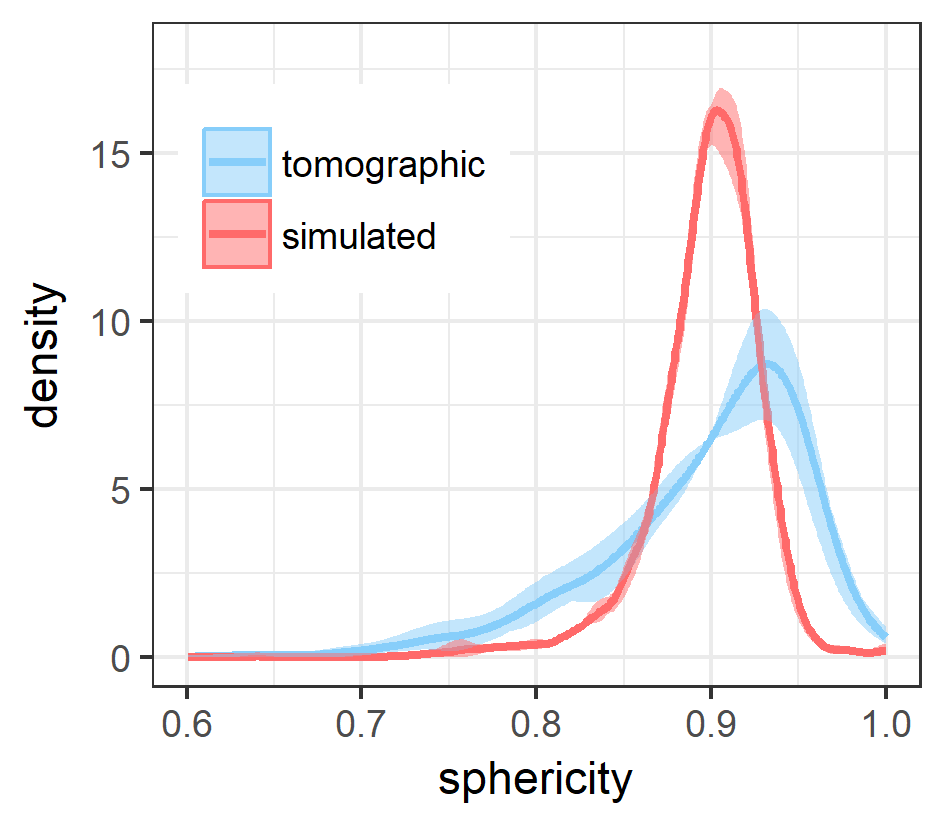}
		\caption{Scenario \textsc{A1}}
		\label{subfig:sphericityPartA1}
	\end{subfigure}
	\begin{subfigure}[c]{0.32\textwidth}
		\includegraphics[width=\textwidth]{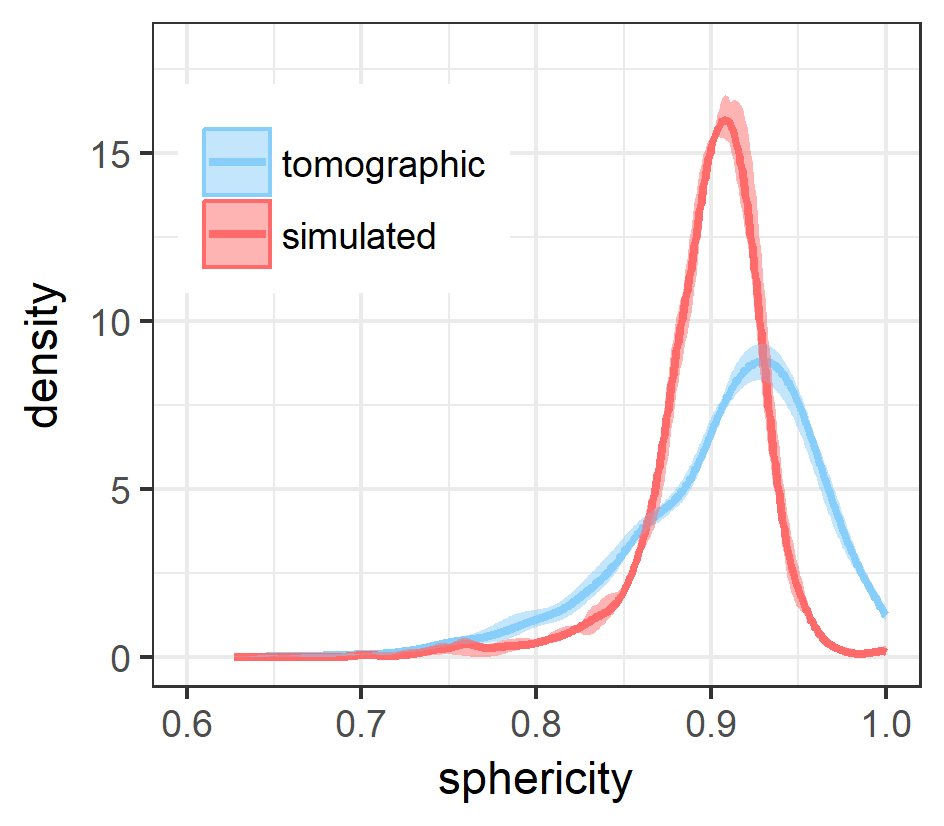}
		\caption{Scenario \textsc{A2}}
		\label{subfig:sphericityPartA2}
	\end{subfigure}
	\caption{Distribution of sphericity of particles.}
	\label{fig:sphericityPart}
\end{figure}

Last but not least, we investigate the sphericity of particles. Nearly spherical-shaped particles are typical for cathodes, but under the given contact conditions in the model it is hard to realize particles of such a shape. Table~\ref{tab:meanSphericity} gives the mean sphericity of particles for each scenario, whereas Figure~\ref{fig:sphericityPart} depicts the corresponding distributions of the sphericity. For each scenario, the mean sphericity of particles from simulated data is very close to that of particles from corresponding tomographic data. Just the distributions of the sphericity slightly differ from each other, because for simulated data the distributions are narrower (i.e., have smaller variances) and do not reach as many values very close to 1 as in tomographic data. Nevertheless, we are able to create particles having acceptable sphericity values as a direct consequence of suitable polytopes for particles and the dynamic $\text{parameter}~L$.
\begin{table}[!ht]
	\centering
	\begin{tabular}{r|c|c|c}
		\hline
		& \multicolumn{3}{c}{mean sphericity of particles}\\
		& \textsc{P} & \textsc{A1} & \textsc{A2}\\
		\hline
		tomographic & 0.8975 & 0.8964 & 0.9071\\
		simulated & 0.902 & 0.8974 & 0.8978\\
		relative error & 0.5\% & 0.12\% & 1.03\%\\
		\hline
	\end{tabular}
	\caption{Mean sphericity of particles.}
	\label{tab:meanSphericity}
\end{table}

%\textcolor{blue}{possible characteristics: pictures of realizations, volume fraction, SSA, nn, pair correlation, (mean) coordination number and zero coord. number, edge length (graph -> distance probability should be related to marks!!!), chord length distributions, sphericity characteristics!!!, intensity, Laguerre sizes (before background points), number of Laguerre facets, surface of Laguerre facets, matchingAreaParticles (estimated area between a particle pair), spherical contact distribution from particles, ...}

%\section{Electrochemical discussion and interpretation of extracted characteristics with respect to the electrical aging scenarios}\label{sec:discussion}
\section{Discussion of changes in microstructure caused by cyclical aging}\label{sec:discussion}

As Section~\ref{sec:valid} showed significant differences between scenarios \textsc{P}, \textsc{A1} and \textsc{A2} with respect to several microstructure characteristics, we will now discuss and interpret these results in detail.

First of all note that electrical aging affects all parts of a Li-ion battery cell, but especially the positive (cathode) and negative (anode) active material. Moreover, electrical aging of scenarios \textsc{A1} and \textsc{A2} does affect anode and cathode in a similar way, so that differences in cell capacity as described in Section~\ref{subsec:material_prepara} are valid indicators for the decrease of performance and the changes in microstructure of anodes and cathodes. In particular, it is well-known that massive decrease of anode capacity is caused by cycling and storage in high battery cell state of charge (SOC). This results in lower electrochemical potentials which lead to degradations like increase of solid electrolyte interface (SEI), deposition of metallic lithium on graphite particles and particle cracking~\cite{vetter.2005}.

In the present paper, cyclic profiles were designed in a way that very high and very low SOCs were reduced to a minimum to avoid the aging mechanisms stated above; in detail, the hybrid-electric-vehicle profile (\textsc{A1}) was cycled in a SOC-range between 70\% and 40\% and the electric-vehicle profile (\textsc{A2}) was set to a range between 90\% and 5\%.
Nevertheless, there are significant differences between the microstructures of scenarios \textsc{P}, \textsc{A1} and \textsc{A2}. Note that in general the morphological characteristics computed for aging scenario \textsc{A1} are often closer to corresponding characteristics of pristine material than those obtained for scenario \textsc{A2}, which suggests that there is evidence for larger mechanical aging stress on cathode side using scenario \textsc{A2}. A common descriptor of this phenomenon is the volume fraction of active particles which is defined as the (expected) volume of active particles per unit volume of the analyzed material. Table~\ref{tab:volSSA} shows the extracted volume fractions for tomographic measurements which nicely mesh with the measured cell capacities from Section~\ref{subsec:material_prepara}. The increase of volume fraction for the cycled samples can be explained by mechanical stress caused by electrical load which leads to a densification of particle systems~\cite{hausbrand.2015}. %, klinsmann.2016, wolfenstine.1999}. %Due to particle-cracking, which, in this case, mainly means a collapsing of the particle system (matrix), the electrical connectivity decreases (which causes the decrease of cathode capacity) and broken parts of the particle matrix fill pores.
Cycling of sample \textsc{A2} (higher depth of discharge (DOD)), causes higher mechanical stress which can be validated by  the highest decrease of capacity as well as the highest volume fraction of active particles observed. 
Furthermore, one can see effects of mechanical aging stress by having a closer look at the continuous pore size distribution (see Figure~\ref{fig:cpsd}). The tomographic data of pristine (Figure~\ref{subfig:cpsdP}) and hybrid-electric-vehicle aged materials (Figure~\ref{subfig:cpsdA1}) show that there is a significant higher amount of larger pores (e.g., pore radii greater $\unit[6]{\mu m}$) available than one can see in Figure~\ref{subfig:cpsdA2} for electric-vehicle aged material. Also the geodesic tortuosity (see Figure~\ref{fig:geodTort}) reflects the phenomenon of densification of the particle systems, which the decreasing occurrence of tortuosity values close to 1 and the (slightly) increasing mean tortuosity show in Figures~\ref{subfig:geodTortP}\hyp{}\ref{subfig:geodTortA2}. Regarding the coordination numbers, we can recognize a shift of their histograms to the right from scenario \textsc{P} via \textsc{A1} to \textsc{A2} (see Figures~\ref{subfig:coordNumP}\hyp{}\ref{subfig:coordNumA2}). This can again be explained by densification mechanisms.

However, most likely, these differences between the microstructures of scenarios \textsc{P}, \textsc{A1} and \textsc{A2} do not yet fully explain the decrease of (electrical) capacity which has been observed for these scenarios, see Section~\ref{subsec:material_prepara}. We thus suppose that there might be further structural degradation phenomena, e.g. regarding the inner structure of the porous active particles, which can not be visualized by the tomographic X-ray imaging technique considered in the present paper.
%\subsection{Discussion of modeling results \textcolor{red}{?? evtl gemeinsam ??}}\label{subsec:discuss2}
%
%What we now want to do is also a brief discussion of the results regarding the model and some of its parameters.
%
%As already mentioned in Section~\ref{subsec:discuss1}, mechanical stress caused by electrical load leads to higher volume fractions and decreasing pores for the aging scenarios \textsc{A2} and especially \textsc{A1}. Whereas between scenario \textsc{P} and \textsc{A2} there are just slight differences in all model parameters, in contrast scenario \textsc{A2} significantly differs in some model parameters. The meant model parameters are the intensity of particles $\lambda$

\section{Summary and outlook}\label{sec:sum_out}

Our intention to enhance the recently developed models for the microstructure of anodes was successfully accomplished and an accordingly adapted model for the microstructure of cathodes could be achieved. The basic concepts of the stochastic 3D model for cathodes are in principle those of the anode models, but several structural differences between anode and cathode materials required modifications of the previous models. This includes the locally occurring large pores in the considered cathodes and the generally low volume fraction of the particle phase. Therefore, a random marked point pattern which particularly models such large pores was incorporated into the modeling approach. By means of a Laguerre tessellation it induces empty pore polytopes which form the desired large pores. Furthermore, the connectivity graph which indicates the particles being connected was slightly modified. For example, the graph does not have to be fully connected anymore. But the most important modifications with respect to the previous anode models concern the shapes of the cathode particles. In contrast to the graphite anode particles, the metal oxide cathode particles exhibit nearly spherical shapes, where especially the larger particles seem to be almost spherical-shaped. For this purpose, two main modifications were made and now the particles in structures simulated by the model exhibit suitable (nearly spherical) shapes. The first modification concerned the Laguerre polytopes into which particles are placed. They initially were too large and had shapes which made it impossible to place nearly spherical-shaped particles into. Therefore, by a rather sophisticated procedure further generators for empty pore polytopes were determined. These additional pore polytopes then shrink the polytopes for particles and simultaneously give them a (more spherical) shape which makes it easier to place nearly spherical-shaped particles into. The second modification concerns the particles themselves and their representation as truncated spherical harmonics series expansions. In the previous anode models, the spherical harmonics series expansion is truncated at a fixed parameter $L$, but now this $L$ is dynamically coupled for each particle to its number of connections given by the connectivity graph. Creating particles this way has two advantages. Particles which have to fulfill a smaller number of connections are less restricted by these contact conditions and have reasonable shapes also for small $L$. The choice of large $L$ is necessary if the particle has to fulfill a large number of connections. Then, a large $L$ gives this particle "more degrees of freedom" such that the particle is able to take a nearly spherical shape.

Finally, the enhanced stochastic 3D modeling approach developed in the present paper can generate virtual microstructures of Li-ion cathodes which resemble their microstructures reconstructed by synchrotron tomography. The goodness of fit between simulated virtual structures and the tomographic ones was confirmed by model validation based on several morphological image characteristics. Moreover, through the alteration of model parameters it was possible to calibrate the model not only to a pristine but additionally to two cyclically aged cathodes.

This latter aspect shows that our cathode model is as flexible as the previously developed anode models and therefore it could also be used for virtual materials testing. That is, the model is able to generate further virtual, not yet manufactured cathode morphologies with optimized performance properties. To investigate cathode morphologies regarding their performance, the simulated structures can then be used as an input for spatially resolved electrochemical simulation models, see~\cite{hein.2016, best.2014}.
%In cooperation with some partners this was already done for virtual anode structures. In combination with the virtual cathode structures it now allows to do electrochemical simulations on a full battery cell.

%In another context, it is scheduled to virtually generate microstructures of differently calendered cathode material by means of this modeling approach.

\section*{Acknowledgements}

This work was partially funded by BMBF under Grant No. 05M13VUA and 03XP0073E.
We are grateful to Matthias Neumann for suggesting the post-segmentation step considered in Section \ref{subsec:data_prepro}.

\section*{References}
\bibliographystyle{unsrt}
%\bibliography{C:/Users/KlausK\string~1/Documents/Promotion/Paper/Bibliography/Literatur}
\bibliography{Literatur}

\end{document}